\documentclass[12pt,a4paper]{article}
\usepackage{jheppub}

\usepackage{graphicx, epsfig}
\usepackage{color}
\usepackage{mathrsfs}
\usepackage{amsmath}
\usepackage{amssymb}

\newcommand{\order}{\mathcal{O}}
\newcommand{\Vol}{\mathcal{V}}
\newcommand{\Kbar}{\bar{K}}
\newcommand{\Pex}{\mathbb{CP}^4_{11169}[18]}
\newcommand{\be}{\begin{equation}}
\newcommand{\ee}{\end{equation}}
\newcommand{\bea}{\begin{eqnarray}}
\newcommand{\eea}{\end{eqnarray}}
\newcommand{\barr}{\begin{array}}
\newcommand{\earr}{\end{array}}

\def\beq{\begin{equation}}
\def\eeq{\end{equation}}
\def\be{\begin{equation}}
\def\ee{\end{equation}}
\def\bea{\begin{eqnarray}}
\def\eea{\end{eqnarray}}

\title{Building an explicit de Sitter}
\author[a,b]{Jan~Louis,} \author[a]{Markus~Rummel,} \author[a]{Roberto~Valandro} \author[c]{and Alexander~Westphal}

\affiliation[a]{II. Institut f\"ur Theoretische Physik der Universit\"at Hamburg, D-22761 Hamburg, Germany}
\affiliation[b]{Zentrum f\"ur Mathematische Physik, Universit\"at Hamburg, D-22761 Hamburg, Germany}
\affiliation[c]{Deutsches Elektronen-Synchrotron DESY, Theory Group, D-22603 Hamburg, Germany}
\emailAdd{jan.louis@desy.de}\emailAdd{markus.rummel@desy.de}\emailAdd{roberto.valandro@desy.de}\emailAdd{alexander.westphal@desy.de}
\abstract{We construct an explicit example of a de Sitter vacuum in type IIB string theory that realizes the proposal of K\"ahler uplifting. As the large volume limit in this method depends on the rank of the largest condensing gauge group we carry out a scan of gauge group ranks over the Kreuzer-Skarke set of toric Calabi-Yau threefolds. We find large numbers of models with the largest gauge group factor easily exceeding a rank of one hundred. We construct a global model with K\"ahler uplifting on a two-parameter model on $\mathbb{CP}^4_{11169}$, by an explicit analysis from both the type IIB and F-theory point of view. The explicitness of the construction lies in the realization of a D7 brane configuration, gauge flux and RR and NS flux choices, such that all known consistency conditions are met and the geometric moduli are stabilized in a metastable de Sitter vacuum with spontaneous GUT scale supersymmetry breaking driven by an F-term of the K\"ahler moduli.}

\begin{document}
\begin{flushright}
DESY 12-146\\
ZMP-HH/12-16\\[0.7cm]
\end{flushright}
\maketitle
\setlength{\parskip}{0.2cm} 

\newpage
\section{Introduction \& Motivation}

String theory is a candidate for a fundamental theory of
nature since it has the capacity to describe chiral 
matter fermions with non-Abelian gauge interactions
within a consistent theory of quantum gravity.
However, at a more detailed level it turns out to be difficult
to clearly identify the Standard Model in an expanding universe as one of the
possible backgrounds.
Part of the problem arises from the poor conceptual understanding of
string theory
which still is largely based on perturbative formulations.
As a consequence there is an abundance of perturbatively consistent 
backgrounds -- each with a generically high-dimensional moduli space. 

This moduli space is particularly apparent if one views the string backgrounds
geometrically, that is as compactifications of a ten-dimensional (10D)
space-time on some compact six-dimensional manifold.
Imposing 4D ${\cal N}=1$ space-time supersymmetry for
phenomenological reasons singles out a specific class of six-manifold
which include  Calabi-Yau (CY) manifolds.
These manifolds have a large number of non-trivial deformations
(moduli) associated with their volume and shape.
In a low energy effective description the moduli
correspond to 4D massless scalar fields which are flat directions 
of the scalar potential. Their stabilization 
has been a long-standing problem but in recent years significant
progress has been made at least for certain classes of compactifications.\footnote{For recent reviews of flux compactifications and their associated
  uplifts to dS, and a much more comprehensive bibliography, see
  e.g.~\cite{Grana:2005jc,Douglas:2006es,Blumenhagen:2006ci,McAllister:2007bg}.} 

The mechanism relevant for this paper uses quantized vacuum
expectation values (VEVs) for $p$-form gauge field strengths of
type IIB string theory \cite{Dasgupta:1999ss,Giddings:2001yu}. These fluxes generate a
scalar potential for a large fraction of the typically ${\cal O}(100)$
moduli
and potentially stabilizes them in a local minimum. The remaining
moduli are then fixed by a combination of non-perturbative effects
(such as gaugino condensation of 4D ${\cal N}=1$ gauge theories living
on D-branes)~\cite{Kachru:2003aw}, a combination of perturbative and
non-perturbative effects
\cite{Balasubramanian:2005zx}, 
or an interplay of perturbative effects and negative curvature of
the internal space alone~\cite{Silverstein:2007ac,Caviezel:2008ik,Haque:2008jz,Caviezel:2008tf,Flauger:2008ad,Polchinski:2009ch,Dong:2010pm,Dong:2011uf}. 

The mechanism for moduli stabilization can simultaneously break
supersymmetry (SUSY) spontaneously 
by generating non-vanishing $F$- and/or $D$-terms \cite{Balasubramanian:2004uy,Burgess:2003ic}. Alternatively, SUSY breaking
can be achieved by inserting an additional
quasi-explicit source, such as an anti-brane in a
warped region \cite{Kachru:2003aw}. The vacuum energy of such fully stabilized
compactifications with SUSY breaking can be both positive and
negative, leading to a description  
of de Sitter (dS) space as metastable vacua of compactified string theory~\cite{Kachru:2003aw,Burgess:2003ic,Balasubramanian:2004uy,Balasubramanian:2005zx,Intriligator:2006dd,Lebedev:2006qq,Haack:2006cy,Cremades:2007ig,Krippendorf:2009zza,Rummel:2011cd,Cicoli:2012fh,Cicoli:2012vw}.
The number of these dS vacua is exponentially large due to large
number of topologically distinct fluxes necessary for moduli
stabilization in the first place. In 
type IIB string theory compactified on a warped 6D Calabi-Yau
manifold there are typically 
${\cal O}(100)$ complex structure or shape moduli associated to the
three-dimensional topologically non-trivial
subspaces (three-cycles) of the Calabi-Yau. On each three-cycle
a flux can be turned on and thus 
one has ${\cal O}(100)$ fluxes to choose for stabilizing all the
complex structure moduli. For, say, 10 available flux quanta per
three-cycle 
this yields ${\cal O}(10^{100})$ isolated potential dS vacua~\cite{Bousso:2000xa,Feng:2000if,Denef:2004ze}. This
exponentially large number of dS vacua with a flat number density
distribution of the vacuum energy is often called the `landscape' of
string vacua. It is coupled to the populating processes of Coleman-deLuccia 
tunneling and eternal inflation, which together realize space-time regions filled with each
dS vacuum infinitely often. 
As a consequence  Weinberg's anthropic argument for the smallness of
the present-day cosmological constant can be realized in string
theory.
Hence the landscape of dS vacua gave string theory the ability to accommodate recent data from observational cosmology which demonstrated late-time accelerated expansion of our visible Universe, consistent with an extremely small positive cosmological constant $\Lambda\sim 10^{-122}\,M_{\rm P}^{4}$.\footnote{The flatness of the vacuum energy distribution on the landscape has recently been questioned by studies using random matrix techniques in general~\cite{Marsh:2011aa,Chen:2011ac,Bachlechner:2012at}, and a statistical analysis of combined input parameter distributions in the context of K\"ahler uplifting~\cite{Sumitomo:2012wa,3rdUTQUEST2012JAPAN}.}

For the purpose of this work we will restrict ourselves to type IIB warped Calabi-Yau orientifold compactifications with three-form flux which arise as a specific weak coupling limit (Sen limit) of F-theory on elliptically fibered CY fourfolds~\cite{Vafa:1996xn,Sen:1997gv}. This will allow us to use the known techniques for constructing CYs, calculate their topological data and derive the 4D effective theory. Supersymmetry breaking and lifting the AdS vacuum to dS typically is the least reliable step. Therefore, our goal is to construct explicit global models in the above type IIB context, which exhibit the dynamics of K\"ahler uplifting.\footnote{In this work, we are not addressing the question of constructing a standard model like sector in combination with moduli stabilization, as it has been recently achieved in~\cite{Cicoli:2011qg,Cicoli:2012vw}. Here, we simply want to avert these complicating features for the sake of moduli stabilization in a stable de Sitter vacuum.} 
In such models the interplay of gaugino condensation on D7-branes and the leading ${\cal O}(\alpha'^{3})$-correction of the K\"ahler potential fix the K\"ahler moduli in a SUSY breaking minimum, after three-form flux has supersymmetrically stabilized the complex structure moduli and the axio-dilaton. The vacuum energy of this minimum can be dialed from AdS to dS by adjusting the fluxes appearing in the superpotential. 
Both SUSY breaking and lifting to dS are driven by an F-term of
the K\"ahler moduli sector which is induced by the presence of the
$\alpha'$-correction~\cite{Balasubramanian:2004uy,Westphal:2006tn,Rummel:2011cd}. The dS uplift is therefore spontaneous and
arises 
from the geometric closed string moduli. This is the motivation for
trying to construct a fully explicit consistent global model including 
a choice for the flux.
Such a model yields an example for a 4D dS space in string theory which
is explicit within the limit of the currently available knowledge.

In this work we  discuss constructions of  K\"ahler uplifted
dS vacua on compact Calabi-Yau manifolds. We perform this analysis
both from perspective of F-theory and its weak coupling limit of
type IIB string theory compactified on warped Calabi-Yau
orientifolds. 
F-theory arises from the observation that the type IIB axio-dilaton can vary over the compactification manifold $B_3$. 
One then interprets the axio-dilaton as the complex structure modulus of an elliptic curve fibered over the threefold $B_3$, realizing a complex four-dimensional manifold that
takes the form of an elliptically fibered Calabi-Yau fourfold in
order to ensure ${\cal N}=1$ supersymmetry in
the 4D effective theory. 
The attraction 
of F-theory is due to the geometrization of the complete
non-perturbative super Yang-Mills (SYM) dynamics of stacks of 7-branes
wrapping four-cycles 
in type IIB in terms of
resolvable ADE type singularities in the fourfold. Of particular
interest here are F-theory compactifications on elliptically fibered
fourfolds which admit a global weak coupling limit (Sen limit), where
the axio-dilaton goes to weak string coupling and becomes approximately constant
everywhere on the threefold base of the fibration. In this limit, 
the set of 7-branes located at the locus where the fiber degenerates,  can be described entirely in terms of 
perturbative O7-planes and D7-branes. 

The analysis in~\cite{Rummel:2011cd} looked at the effective dynamics
of moduli stabilization via K\"ahler uplifting. A class of possible
examples consists of a Calabi-Yau threefold with an expression for the
volume $\Vol$ given by the sum of one large four-cycle modulus and a
collection of smaller so called 
blow-up four-cycle moduli. This resembles the structure of a swiss
cheese with its overall volume given mainly by just the size of the
enclosing cycle and many tiny holes (the blow-up four-cycles).  
We know in addition that in K\"ahler uplifted dS vacua the volume $\Vol$ of the type IIB Calabi-Yau scales with the rank $N$ of the condensing gauge group as $\Vol \propto  N^{3/2}$. Moreover, the scale of K\"ahler moduli stabilization and thus the resulting K\"ahler moduli masses are suppressed by an additional ${\cal O}(1/\Vol)$ compared to the scale of the flux-induced complex structure moduli stabilization. 
These features lead us to search for condensing
gauge groups  with a large rank  which induce a large volume.

We will consider models which can be easily uplifted to F-theory compactified on elliptically fibered CY fourfolds that are hypersurfaces in 
an ambient toric variety~\cite{Collinucci:2008zs,Collinucci:2009uh,Blumenhagen:2009up}.  
In addition, we insist on the existence of a smooth Sen limit, as we
use the leading $\alpha'$ correction which is not understood in
F-theory in general. The type IIB Calabi-Yau threefold is then
the double cover of the F-theory base manifold $B_3$. It is
described by an equation such as $\xi^2=h$, where the orientifold
involution is realized by $\xi\mapsto -\xi$. Here $\xi$ denotes one of
the holomorphic complex projective coordinates of the ambient toric
variety  where the CY threefold lives, with the orientifold plane sitting at $\xi=0$. 
The data describing the D7-brane stacks and the orientifold planes in
F-theory via ADE singularities can be specified in terms of sections
of holomorphic line bundles and the corresponding homology classes of the
associated divisors. 
The D7-brane tadpole forces the D7-brane to wrap cycles whose homology classes 
(denoted by $[D7]$) sum up to $[D7]=8[O7]=8[\{\xi=0\}]$ in the type IIB CY threefold. To obtain a large gauge group one generically has to wrap many branes on certain divisors. This is only possible if the coefficients of the orientifold class $[O7]=\sum_i c_i [D_i]$ are large (here the $D_i$ denote a complete set of divisor four-cycles which in turn form a base of the 2nd Dolbeault cohomology group~$H^{1,1}$).
Increasing the homology class $[O7]$  
introduces singularities in the base manifold which have to be resolved. Moreover, the Calabi-Yau threefold hypersurface in the weak coupling limit should be free of singularities as well, and therefore the orientifold planes should not intersect each other. This in general turns out to be a severe constraint when one tries to increase the class of the orientifold by choosing  the weights defining the toric variety appropriately.

Finally, we have to check that all used divisors are rigid such that
gaugino condensation does contribute to the superpotential.
In this context, the role of gauge flux is crucial: On the
one hand, a suitable choice can `rigidify' a divisor by fixing some
of its deformation moduli \cite{Martucci:2006ij,Bianchi:2011qh} (see
\cite{Lust:2005bd,Braun:2008pz,Braun:2011zm} for discussions in the
F-theory context).
On the other hand, switching on gauge flux can generate additional zero modes in the form of chiral matter (especially at the intersection of branes) 
which  forbids the contribution of gaugino condensation in the superpotential.
Moreover, the presence of fluxes can be required by the necessity of 
canceling the Freed-Witten anomaly~\cite{Minasian:1997mm,Freed:1999vc}.

As mentioned before, the dynamics of K\"ahler uplifting was
demonstrated so far on `swiss cheese' type Calabi-Yau threefolds. The
cheese with its one big bounding cycle and its many tiny holes implies
a certain form of the volume. Let us assume a CY with a set
of divisor four-cycles $D_i$ whose (real) volumes we denote by
$\Vol_i$. Then the volume of the a swiss cheese CY is defined by $\Vol
\sim \Vol_1^{3/2} - \sum_i\Vol_i^{3/2}$ or as
$\Vol \sim (\Vol_1 + \sum_i\Vol_i)^{3/2} -\sum_i
\Vol_i^{3/2}$ for an approximately swiss cheese CY.
In this situation one can manufacture a large overall volume by enforcing a large gauge group rank on the corresponding divisor $D_1$ which in turn leads to
a large $\Vol_1$. 
A potential complication may then arise as a high number of branes on $D_1$ typically enforces singularities on other divisors which might yield the overall volume small even though $\Vol_1$ is large.  

At the end, we combine the K\"ahler moduli stabilization with an explicit dilaton and complex
structure moduli stabilization via RR and NS fluxes. We
discuss a specific hypersurface in the weighted projective space
$\mathbb{CP}^4_{11169}$ as a concrete example where the whole program can be executed.
In this case, discrete symmetries of the complex structure moduli space  and a specific choice of the three-form fluxes
allow us to fix all the complex structure explicitly along the lines of~\cite{Giryavets:2003vd}.
We check that this choice of flux
results in values for $W_0$ which are such that
the K\"ahler stabilization leads to 
a metastable dS vacuum. 
Hence, our model constitutes an example for a dS space in string theory which is explicit within the limits of existing knowledge. The only implicitness that remains is the unknown  complex structure moduli dependence of the 1-loop determinant prefactor of the non-perturbative effect. Recent work has shown~\cite{Rummel:2011cd} that for large volume the mass scale of the K\"ahler moduli separates from the scale of the axio-dilaton and the complex structure moduli by one inverse power of the volume. This justifies replacing the complex structure moduli by their VEVs inside the 1-loop determinants, and allows us to parametrize these prefactors as effective constants. Moreover, we can clearly dial the VEVs of the complex structure moduli by availing ourselves of the exponentially large flux discretuum, which easily accounts for a potential mild tuning of the value of the 1-loop determinants.

This paper is organized as follows.
In Section~\ref{lggr_sec} we study the constraints for having 
a gauge group with large  rank by discussing Kreuzer-Skarke models~\cite{Kreuzer:2000xy} and hypersurfaces in toric varieties. For the subclass of threefolds with an elliptic F-theory lift ($\sim 10^{5}$ models) we scan and extract the distribution of the largest-rank gauge group as a function of the number of K\"ahler moduli $h^{1,1}$.
Then we choose to consider $\Pex$ as an explicit example and 
construct large-rank ADE gauge groups on a choice of
two divisors, and analyze the consistency constraints both in the type
IIB weak-coupling limit, and from the F-theory perspective in
sections~\ref{IIBperspectiveEx} and~\ref{FthEx_sec}, respectively.
Section~\ref{CSlargeVolume} reviews the general results for
supersymmetric flux stabilization of the complex structure moduli and
the axio-dilaton. 
In section~\ref{dS_sec} we study the scalar potential that stabilizes
the K\"ahler moduli. We single out a band in the $g_s$ - $W_0$
plane where one finds de Sitter vacua. Here $W_0$ denotes the VEV of
the superpotential which arises from supersymmetric flux stabilization of the complex structure moduli. 
For the explicit model on $\Pex$, we show how to fix explicitely all the complex structure moduli, thanks to a particular symmetry of the moduli space and a special choice of three-form fluxes.
Finally, we check that this choice of flux
results in values for $W_0$ which are such that
the K\"ahler stabilization leads to 
a metastable dS vacuum. 
We conclude and discuss our results in section~\ref{conclusions_sec}. More details of the toric resolution of the $Sp(k)$-singularity can be found in the appendices.
We have kept the 
steps of our calculations 
rather explicit for future reference but also since certain aspects of the 
arguments are often  only implicit in the existent literature.

\section{Constraints on large gauge group rank in the landscape} \label{lggr_sec}

In this section, we discuss generic constraints on obtaining large gauge group gaugino condensation which is a crucial input for the method of K\"ahler uplifting. In the context of non-compact Calabi-Yaus, it was already discussed in~\cite{Gukov:1999ya} that arbitrarily high gauge group ranks are possible. As we will see, the situation in the compact case is more restrictive. We will mostly discuss the perturbative type IIB picture and conclude with some remarks about non-perturbative F-theory models at the end of this section.

Our laboratory will be the landscape of  complex three-dimensional
Calabi-Yau manifolds that are hypersurfaces in toric varieties. These
were classified in~\cite{Kreuzer:2000xy} by constructing all
473,800,776 reflexive polyhedra that exist in four dimensions,
yielding 30,108 distinct Hodge numbers of the corresponding Calabi-Yau
manifolds $X_3$. For simplicity we will study a subset of these, i.e.\ the set of 184,026 maximal polytopes yielding 10,237 distinct Hodge numbers. These can be represented by a weight system of positive integers $n_1,...,n_5$. To each integer $n_i$ one can associate
one of the projective coordinates $\{u_1,...,u_4,\xi\}$ of a four-dimensional toric space:
\begin{equation}
\begin{array}{ccccc}
u_1 & u_2 & u_3 & u_4 & \xi \\
\hline
n_1 & n_2 & n_3 & n_4 & n_5 \\
\end{array}
\qquad \text{with} \qquad 0 < n_1 \leq n_2 \leq n_3 \leq n_4 \leq n_5\,\,.
\label{WSdef}
\end{equation}
The integers $n_i$ determine the scaling equivalence relation the coordinates satisfy:
\begin{equation}\label{equivrelamb}
 (u_1,...,u_4,\xi) \sim (\lambda^{n_1}u_1, ..., \lambda^{n_4}u_4, \lambda^{n_5}\xi) \,, \qquad \mbox{with } \,\, \lambda\in \mathbb{C}^\ast \:.
\end{equation}
The divisors $D_i:\{u_i=0\}$ and $D_\xi:\{\xi=0\}$ are called toric divisors.  
A hypersurface in such toric space is a Calabi-Yau (i.e. its first Chern class vanishes) if the degree of the defining equation is equal to $\sum_i^5 n_i$.

Eq.~\eqref{equivrelamb} defines the  complex four-dimensional
projective space $\mathbb{CP}^4_{n_1 n_2 n_3 n_4 n_5}$. Often it is
useful to think about the weights as defining a gauged linear sigma
model (GLSM)~\cite{Witten:1993yc}. If one of the weights $n_i$ is greater than one, the ambient space is not smooth. This is the case for any toric Calabi-Yau that is not the quintic, which is given by $n_i=1$, $\sum_i^5 n_i=5$. The corresponding singularities have to be resolved if they intersect the Calabi-Yau hypersurface. The resolution process yields additional weights, i.e.\ eq.~\eqref{WSdef} becomes a $k\times (k+5)$ matrix, called the weight matrix, that defines the resolved toric ambient space $X_4^{\rm amb}$. Generically, the greater the $n_i$ in eq.~\eqref{WSdef}, the more lines of weights have to be added to obtain a smooth Calabi-Yau. Often there is more than one choice to resolve the singularities, corresponding to different triangulations of the corresponding polytope. The number of lines of the weight matrix $k$ gives the 
dimension of $H^{1,1}(X_4^{\rm amb},\mathbb{Z})$. Since some divisors of $X_4^{\rm amb}$ might either intersect $X_3$ in two or more disconnected and independent divisors of $X_3$, or even not intersect $X_3$ at all, dim$H^{1,1}(X_4^{\rm amb},\mathbb{Z})$ is not necessarily the same as $h^{1,1}=\text{dim } H^{1,1}(X_3,\mathbb{Z})$. However, increasing $k$ will generically also increases $h^{1,1}$.

To realize an ${\cal N}=1$ supersymmetric compactification of type IIB in four dimensions and to consistently include D-branes and fluxes we introduce O7 orientifold planes in the construction. For simplicity, we only consider orientifold projections $\mathcal{O}=(-1)^F\Omega_p \sigma$ acting via the holomorphic involution 
\begin{equation}\label{OrIvolXi}
 \sigma : \quad \xi \mapsto -\xi\,,
\end{equation}
i.e.\ the sign of the coordinate with the highest weight is reversed. 
We demand
\begin{equation}
n_\xi \equiv n_5 = \sum_{i=1}^4 n_i\,,
\label{FthmodelWS}
\end{equation}
such that the Calabi-Yau hypersurface equation symmetric under \eqref{OrIvolXi} is given by
\begin{equation}
 \xi^2 = P_{(2 \sum_{i}^4 n_i, \dots)}\,.
\label{xi2sumni}
\end{equation}
The dots denote possible additional weights that have to be added to
obtain a threefold free of singularities. Note that
eq.~\eqref{xi2sumni} only holds if $n_\xi = \sum n_i$ also for the
resolution weights which we assume is in many cases possible and which we have verified in various examples. 

Hence, all information of the Calabi-Yau threefold is stored in the weights $n_1,\dots, n_4$ and the chosen triangulation. Moreover, the resolution of the  three dimensional manifold $\mathbb{CP}^3_{n_1 n_2 n_3 n_4}$ is the base $B_3$ of the elliptically fibered fourfold that realizes the uplift of the type IIB model to F-theory. For this reason, models fulfilling eq.~\eqref{FthmodelWS}, are named models of the `F-theory type'. These are 97,036 weight systems leading to 7,602 distinct pairs of Hodge numbers. The first Chern class of $B_3$ defines a non-trivial line bundle, the anti-canonical bundle $\Kbar$,  with $\Kbar=c_1(B_3)$ (in this paper we use the same symbol to denote the line bundle and its corresponding divisor class). Due to eq.~\eqref{FthmodelWS} the homology class of the O7-plane at $\xi=0$ is given as $[O7]=\Kbar$.

\subsection{D7-branes from the IIB perspective} \label{D7IIBgen_sec}

Now, we discuss the inclusion of D7-branes from the IIB perspective. The presence of the O7-plane induces a negative D7 charge of $-8 [O7]$. This has to be compensated by the positive charge of the D7-brane stacks $[D7]$ 
($[D7]$ is the homology class of the surface wrapped by the D7-brane configuration). In other words, since $[O7]=\Kbar$, $[D7]$ has to be given by the vanishing locus of a section of $\Kbar^8$ to saturate the D7 tadpole. More specifically, it was found in~\cite{Collinucci:2008pf} that for a single invariant D7-brane saturating the D7 tadpole cancellation condition, its world volume is given by the (non-generic) polynomial equation
\begin{equation}\label{WhitneyBr}
 \eta^2-\xi^2\chi = 0\,,
\end{equation}
with $\eta$ and $\chi$ sections of $\Kbar^4$ and $\Kbar^6$, respectively. (For practical purposes, $\eta$ and $\chi$ can be seen as polynomials in the complex coordinates $u_i$ of the resolved base manifold $B_3$.) 
This brane can be understood as the result of the recombination of one standard D7-brane wrapping the surface $\eta-\xi\psi=0$ with its orientifold image, wrapping $\eta+\xi\psi=0$. In fact, such a brane configuration is described by the vanishing locus of the factorized polynomial $(\eta-\xi\psi)(\eta+\xi\psi)=\eta^2-\xi^2\psi^2$.  By adding to this polynomial the term $\xi^2(\psi^2-\chi)$, i.e. by recombining the two factors, one obtains the equation \eqref{WhitneyBr}. The resulting recombined invariant D7-brane is called in literature `Whitney brane', as it has the singular shape of the so called Whitney umbrella \cite{Collinucci:2008pf}.

 For non-generic forms of the polynomials $\eta$ an $\chi$, 
the Whitney brane can split into different stacks. In particular a stack of
$2 N_i$ branes wrapping the invariant toric divisor $D_i:\{u_i=0\}$ manifests itself via the factorization
\begin{equation}\label{etachiSpN}
 \eta = u_i^{N_i} \tilde \eta\,,\qquad \qquad \quad \chi = u_i^{2 N_i} \tilde \chi\,,
\end{equation}
such that eq.~\eqref{WhitneyBr} becomes
\begin{equation}
u_i^{2 N_i} \left( \tilde \eta^2-\xi^2 \tilde \chi \right) = 0\,,
\label{branefacgen}
\end{equation}
where on the invariant divisor at $u_i=0$ there is an $Sp(N_i)$ stack and $\tilde \eta^2-\xi^2 \tilde \chi$ describes a Whitney brane of lower degree. Since the Whitney brane has always to be described by a holomorphic equation, $N_i$ cannot be made arbitrarily large.

For $u_{i=1,\dots,4}$ we can be more specific. Eq.~\eqref{branefacgen} becomes
\begin{equation}
u_i^{2 N_i} \left( \tilde \eta_{(4 n_\xi - n_i N_i,\dots)}^2-\xi^2 \tilde \chi_{(6 n_\xi - 2 n_i N_i,\dots)} \right) = 0\,,
\label{branefac14}
\end{equation}
where the dots denote the degrees that are imposed via the weight system of the resolved ambient space $X_4^{\rm amb}$. If the degree in the first scaling is the most restrictive we obtain the strongest bound from the holomorphicity of $\tilde \chi$, i.e.\
\begin{equation}
 N_i \leq 3 \frac{n_\xi}{n_i}\,.
\label{indicatorIIB}
\end{equation}
Due to the ordering of the $n_i$, eq.~\eqref{WSdef}, we expect to be able to put the largest number of branes on the divisor $D_1$ and the constraining quantity is  
the largest integer $N_{lg}$ that is smaller than $3 n_\xi / n_1$. $N_{lg}$ will serve as our large gauge group indicator in the following.

Before we proceed, let us make a few comments on the choice of the large gauge group indicator that were in part already addressed in the introduction:
\begin{itemize}
\item For $n_1 = 1$, the first column of the $k\times (k+5)$ weight matrix describing the resolved ambient space $X_4^{\rm amb}$ is always given by $(1,0,\dots,0)^T$ and hence $N_{lg}$ is always the limiting quantity. However, for $n_1 > 1$ the first column of the weight matrix has to contain additional non-zero entries smaller than $n_1$ to resolve the singularities. The holomorphicity of the Whitney brane equation in the corresponding degree could in principle be more restraining than $N_{lg}$. Even if this would be the case we still expect $N_{lg}$ to give a right estimate since in the $\order(10)$ examples where we have computed the resolved weight matrix, using \textit{PALP}~\cite{Kreuzer:2002uu,Braun:2011ik,Braun:2012vh}, it was always the most restrictive.
\item The type of gauge group enforced by the $2 N_1$ branes depends on the geometry of the O7-plane and the gauge flux. If the divisor is invariant under $\xi \mapsto - \xi$ and transverse to the O7-plane, we have an $Sp(N_1)$ gauge group that can be broken by gauge flux to $SU(N_1)$. The Coxeter numbers are $N_1+1$ for $Sp(N_1)$ and $N_1$ for  $SU(N_1)$. If the divisor lies on the orientifold plane the gauge group is $SO(2 N_1)$ with Coxeter number $2N_1-2$.
\item To check if the brane-stack contributes to the superpotential in a suitable way for the method of K\"ahler uplifting one has to fulfill additional constraints. First of all, in order to have a pure SYM theory that undergoes gaugino condensation, possible light matter fields must be forbidden. A sufficient condition is that 
the wrapped divisor is rigid and that the brane intersections and world volume should not produce additional chiral zero-modes.  
Note that with growing rank of the weight matrix, $D_1$ tends to be rigid since it typically cannot be deformed into other toric divisors. Furthermore, the volume form of the threefold has to be of the approximately swiss cheese type. Finally, the factorization in eq.~\eqref{branefac14} should not force a further factorization of the remaining Whitney brane in toric divisors which enter the volume form with a negative sign. In fact, since gaugino condensation forces the volumes of these divisors to be large, this would make the overall volume small. This does not necessarily have to be a problem since in the approximately swiss cheese type the enforced brane stacks on other toric divisors might also increase the overall volume.
\end{itemize}
We mention these points to make it clear that the indicator $N_{lg}$ only serves as an easily computable estimate for the largest gauge group rank one can obtain in a threefold of the F-theory type. 
To see if one can stabilize the K\"ahler moduli in a large volume, one has to check the additional constraints case by case. We will do this in Section~\ref{IIBperspectiveEx}, constructing a consistent model of a K\"ahler uplifted de Sitter vacuum.

\subsection{D7-branes from the F-theory perspective}

Let us now discuss the constraints on the large gauge group rank in the perturbative limit of F-theory. This theory is physically equivalent to weakly coupled type IIB, discussed in the previous section. However, the geometric F-theory picture provides a different perspective and a cross check of our results.

Before we discuss the D7-brane setup in F-theory let us set the stage. To obtain an $\mathcal{N}=1$ effective four dimensional effective theory starting from 12-dimensional F-theory we have to compactify on an elliptically fibered Calabi-Yau fourfold. More specifically, the fourfold can be described as a hypersurface in an ambient fivefold which is a $\mathbb{CP}^2_{123}$ fibration over a three dimensional base $B_3$, i.e.\ one introduces three additional complex coordinates coordinates and a scaling relation
\begin{equation}
 (X,Y,Z) \sim (\lambda^2 X,\lambda^3 Y,\lambda Z)\,.
\label{P123scaling}
\end{equation}
As far as the scaling in the classes of the base is concerned $Z$ scales as a section of  the canonical bundle $K$ of $B_3$, in order to ensure the Calabi-Yau condition of the fourfold. The elliptically fibered Calabi-Yau fourfold can be defined by the Weierstrass model
\begin{equation}
Y^2  = X^3 + f\, X Z^4 + g\, Z^6\,,
\label{Weierstrassform}
\end{equation}
with $f$ and $g$ being sections of $\Kbar^4$ and $\Kbar^6$, respectively. However, for the purpose of detecting singularities it is more convenient to bring~\eqref{Weierstrassform} in the Tate form~\cite{Tate:1975,Bershadsky:1996nh}:
\begin{equation}
 Y^2 + a_1 X Y Z + a_3 Y Z^3 = X^3 + a_2 X^2 Z^2 + a_4 X Z^4 + a_6 Z^6\,,
\label{Tateequation}
\end{equation}
where the Tate polynomials $a_i$ are functions of the base coordinates $u_i$ such that they are sections of $\Kbar^i$. The Tate form~\eqref{Tateequation} and the Weierstrass form~\eqref{Weierstrassform} of the defining equation can be related by completing the square and the cube and shifting the $X,Y$ coordinates.  

In F-theory, D7-branes manifest themselves via singularities of the elliptic fibration. To engineer a singularity on a divisor $D_j:\{u_j=0\}$ the Tate polynomials have to factorize as
\begin{equation}
 a_i = u_j^{w_i} a_{i,w_i}\,,
\label{aifacgen}
\end{equation}
with positive integer numbers $w_i$ encoding which kind of singularity
is realized. Since $a_{i,w_i}$ has to be holomorphic, $w_i$ cannot be
made arbitrarily large. For a tabular overview of the possible
resolvable singularities that can arise in such a construction
see~\cite{Donagi:2009ra}. The singularities with Coxeter number larger
than 30 are either of the $Sp$, $SU$ or $SO$ type (see table \ref{SpSUSO}).

\begin{table}[ht!]
\centering
  \begin{tabular}{|c|cccccc|}
  \hline
   & $a_1$ & $a_2$ & $a_3$ & $a_4$ & $a_6$ & $\Delta$\\
  \hline
  $Sp(N)$ &  $0$ & $0$ & $N$ & $N$ & $2 N$ & $2 N$\\
  $SU(2N)$ &  $0$ & $1$ & $N$ & $N$ & $2 N$ & $2 N$\\
  $SU(2N+1)$ &  $0$ & $1$ & $N$ & $N+1$ & $2 N+1$ & $2 N+1$\\
  $SO(4N+1)$ &  $1$ & $1$ & $N$ & $N+1$ & $2 N$ & $2N + 3$\\
  $SO(4N+2)$ &  $1$ & $1$ & $N$ & $N+1$ & $2 N+1$ & $2N + 3$\\
  $SO(4N+3)$ &  $1$ & $1$ & $N+1$ & $N+1$ & $2 N+1$ & $2N + 4$\\
  $SO(4N+4)$ &  $1$ & $1$ & $N+1$ & $N+1$ & $2 N+1$ & $2N + 4$\\ \hline
  \end{tabular}
  \caption{The exponent $w_i$ in \eqref{aifacgen} is given for all $a_i$ and $\Delta$ and for different singularities. The discriminant $\Delta$ depends on the $a_i$ according to eq.~\eqref{Deltabi}.}
  \label{SpSUSO}
\end{table}

We can again analyze the constraints on the maximal gauge group rank in more detail. Since the anti-canonical class of the base $B_3$ is given by $\Kbar=\sum_{i=1}^4D_i$ and $n_\xi=\sum_{i=1}^4 n_i$,
eq.~\eqref{aifacgen} can be written as
\begin{equation}
 a_{(i\,n_\xi,\dots)} = u_j^{w_i} a_{(i\,n_\xi- w_i n_j,\dots )}\,,
\label{aifacWS}
\end{equation}
where the dots once more denote scalings originating from the resolution of the singularities of the original weight system of the base. If one considers the singularities listed in Table~\ref{SpSUSO}, the most severe constraints regarding holomorphicity of eq.~\eqref{aifacWS} come from $a_3$ and $a_6$. A sufficient condition for the $a_i$ to always be holomorphic at least under the first scaling is
\begin{equation}
 N_j \equiv w_3=\frac{w_6}{2}\leq 3 \frac{n_\xi}{n_j}\,,
\end{equation}
which is exactly what we found in
eq.~\eqref{indicatorIIB} in the type IIB picture. Thus, also from the F-theory perspective we
arrive at the large gauge group indicator $N_{lg}$. The caveats discussed in the comments following eq.~\eqref{indicatorIIB} of course also have to be taken into account in the F-theory picture.

So far our F-theory discussion has been for generic values of the string coupling. However, we eventually want to obtain a stable de Sitter vacuum by using the leading $\alpha'$ correction to the K\"ahler potential~\cite{Becker:2002nn} which is only known in perturbative type IIB. 
As long as this correction remains unknown in non-perturbative F-theory, we have to restrict our analysis to Sen's weak coupling limit $g_s\rightarrow 0$~\cite{Sen:1997gv}. 

In the Tate form~\eqref{Tateequation}, the Sen limit \cite{Sen:1997gv} is imposed by the rescalings \cite{Donagi:2009ra}
\begin{equation}
 a_3\mapsto \epsilon\, a_3\,,\qquad a_4\mapsto \epsilon\, a_4\,,\qquad a_6\mapsto \epsilon^2\, a_6\,.
\label{aiSen}
\end{equation}
and $\epsilon\rightarrow 0$. The string coupling is related to the parameter $\epsilon$ by
\begin{equation}
 g_s \sim -\frac{1}{\log|\epsilon|} \rightarrow 0\qquad \text{as}\qquad \epsilon \rightarrow 0\,.
\end{equation}

Completing the square and the cube in eq.~\eqref{Tateequation} gives a relation between $f$, $g$ in the Weierstrass model and the Tate polynomials:
\begin{equation}
 f = -\frac{1}{48}(h^2 - 24 \epsilon \eta)\,,\qquad g=-\frac{1}{864}(-h^3+36 \epsilon h \,\eta - 216 \epsilon^2\chi)\,,
\end{equation}
with
\begin{equation}
 h = a_1^2+4 a_2  \,,\qquad \eta=a_1 a_3 +2 a_4 \,,\qquad \chi = a_3^2 +4 a_6\,,
\end{equation}
where $h,\, \eta$ and $\chi$ are sections of $\Kbar^2,\, \Kbar^4$ and $\Kbar^6$ respectively.
The discriminant locus $\Delta=0$, where the elliptic fiber degenerates, gives the location of the D7/O7-planes. The discriminant is given by 
\begin{equation}
 \Delta = \frac{1}{16}\left( \epsilon^2 h^2 P_{D7} + 8 \epsilon^3 \eta^3 + 27 \epsilon^4 \chi^2 - 9 h \epsilon^3\eta \chi \right)\,  \sim \frac{1}{16} \epsilon^2\, h^2 P_{D7} + \order(\epsilon^3)\:. 
\label{Deltabi}
\end{equation}
where $P_{D7} = -\frac{1}{4}( \eta^2-h \chi)$. We see that in the weak coupling limit $\epsilon\rightarrow 0$, the locus $\Delta=0$ splits into two components. Studying the monodromies of the elliptic fiber around such components, one realizes that at $h=0$ there is an O7-plane, while at $P_{D7}=0$ there is a D7-brane \cite{Sen:1997gv}.
The type IIB Calabi-Yau hypersurface that is a double cover of $B_3$ is given by
\begin{equation}
 X_3:\,\, 0= \xi^2 - h = \xi^2 - (a_1^2 + 4 a_2) \,.
\label{CY3TateSen}
\end{equation}
Using this equation, we see that the relation defining the D7-brane configuration in the Calabi-Yau threefold is
\begin{equation}
 P_{D7}=0 \,, \qquad \mbox{ i.e. } \qquad \eta^2-\xi^2\chi =0\ .
\end{equation}
As said above, a brane wrapping such a surface has the form of a Whitney umbrella
\cite{Collinucci:2008pf} and will be called  a `Whitney brane' in this paper. 

Note that there is no restriction on $f$ and $g$ or the Tate polynomials $a_i$ respectively in the Sen limit. Hence, a singular configuration over a divisor $D_j$ enforced via a factorization $a_i = u_j^{w_i} a_{i,w_i}$ remains intact in the weak coupling limit. This is consistent with the fact that we found the same bound on the gauge group rank in the IIB picture and in the general F-theory picture.

However, one has to make sure that a non-singular Sen limit exists. For example, $SU(N)$ singularities generically introduce conifold singularities in the type IIB Calabi-Yau threefold (located on top of the O7-plane) \cite{Donagi:2009ra}. In case of $SU(N)$ singularities, the set of F-theory bases that lead to a smooth type IIB Calabi-Yau threefold is only a subset of the singular free bases whose elliptic fibrations generate F-theory fourfolds \cite{Krause:2012yh}. 
Therefore, we expect the constraints on the gauge group rank to be less strict in
a generic F-theory compactification. In F-theory
the constraints are that the base is non-singular and that one is not forced to introduce fibre singularities that cannot be resolved according to the Kodaira or Tate classification. Hence, it would be very interesting if one could derive the analogue of the 
perturbative $\alpha'$ correction to the K\"ahler potential in general F-theory, making the construction of truly non-perturbative de Sitter vacua accessible.

\subsection{Maximal gauge group ranks}

In this section, we give the results of our scan for the maximal gauge
group indicator $N_{lg}$ in the 97,036 models of the F-theory type
contained in the classification of~\cite{Kreuzer:2000xy}. We use
\textit{PALP} to calculate the Hodge numbers resulting in 7,602
distinct pairs of Hodge numbers. We refer to this set of weight
systems as the general set. We also gather all weight systems that
lead to the same pair of Hodge numbers and choose as the
representative the weight system with the smallest $N_{lg}$
corresponding to the most conservative estimate for the maximal gauge
group. This set of weight systems is referred to as the conservative
set. We restrict our attention to manifolds with negative Euler number $\chi = 2(h^{1,1} - h^{2,1})$, further reducing the set of weight systems to 8,813 corresponding to 3,040 distinct pairs of $(h^{1,1},h^{2,1})$. We do this since $\chi < 0$ is a necessary condition to apply the method of K\"ahler uplifting.

\begin{figure}[t!]
\centering
\includegraphics[width= \linewidth]{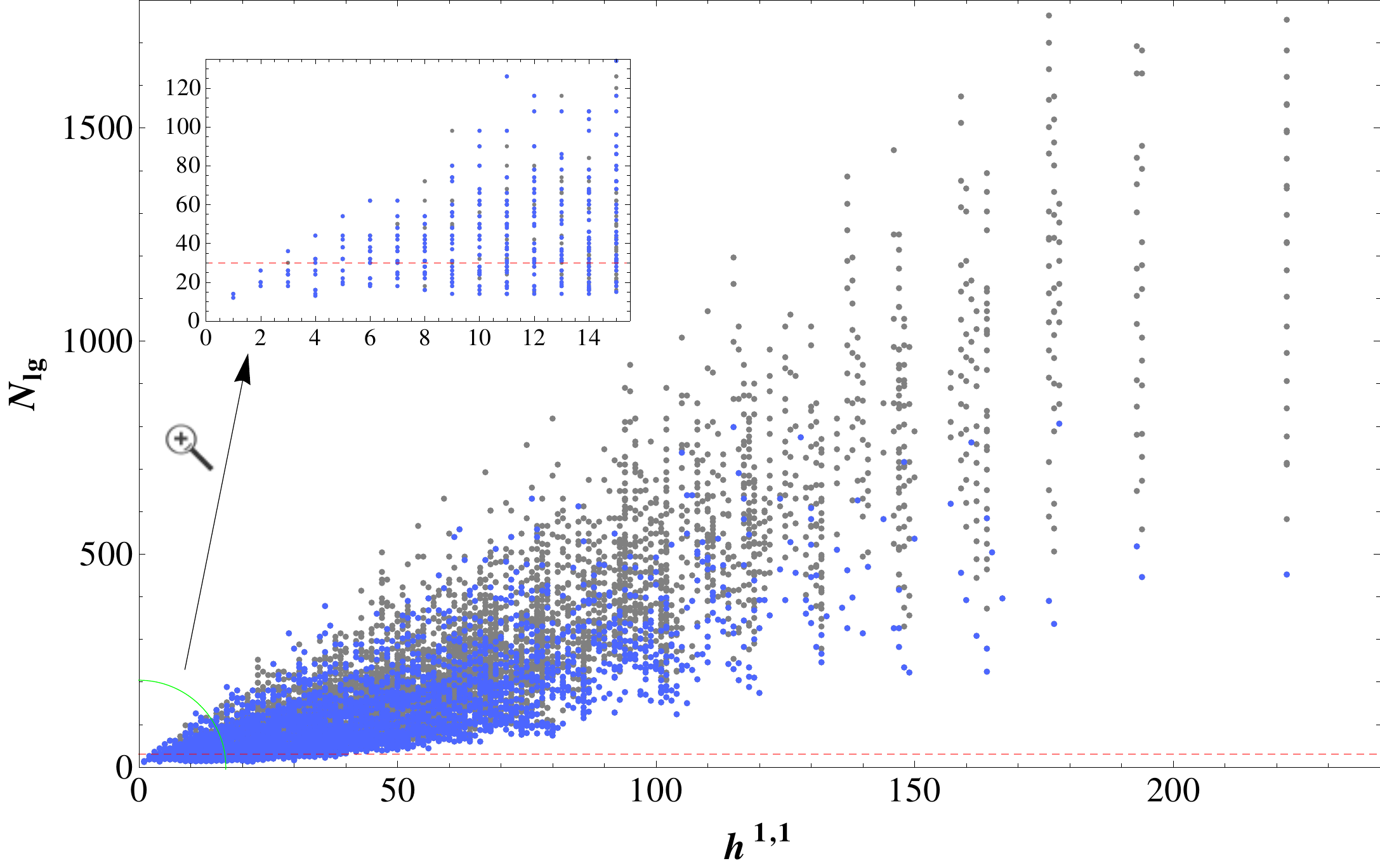}
\caption{The large gauge group indicator $N_{lg}$ as a function of $h^{1,1}$. The grey dots denote the general set of models, while the blue dots denote the conservative set (for explanations see text). The red dashed line denotes the critical gauge group rank for K\"ahler uplifting $N_{lg}^{\text{crit.}} = 30$.}
\label{Xin1vsh11}
\end{figure}

Our results are summarized in Figure~\ref{Xin1vsh11} and
Table~\ref{Xin1vsh11_tab}. The maximal $N_{lg}$ we obtain is 2,330 in
the general set and 806 in the conservative set. The minimal $N_{lg}$
is 12 in both sets corresponding to the base $\mathbb{CP}^3_{1111}$. The
mean $\bar N_{lg}$ we find is 204.5 in the general set and 132.8 in
the conservative set. Generically, the critical value for K\"ahler uplifting to be in large volume regime
($\Vol \gtrsim 100$) is $N_{lg}^{\text{crit.}} = 30$ . Since the actual volume also depends on the
intersection numbers and the stabilized volumes of the divisors other
than $D_1$, $N_{lg}^{\text{crit.}}=30$ can only serve as an estimate for
large volume. The subset of weight systems with $N_{lg} <
N_{lg}^{\text{crit.}}$ is 444 in the general set and 267 in the
conservative set, corresponding to only $5\%$ respectively $9\%$ of
the models where the method of  K\"ahler uplifting is not applicable.

Another important feature we notice is the dependence of $N_{lg}$ on $h^{1,1}$. We see from figure~\ref{Xin1vsh11} that $N_{lg}$ tends to increase with $h^{1,1}$. In other words, if one wants to have very large gauge groups one has to buy this by a rather high number of K\"ahler moduli which of course has the disadvantage of increasing the complexity of the model, especially if it is not swiss cheese.\footnote{For an algorithm to check for the swiss cheese property of a threefold see~\cite{Gray:2012jy}.} The tendency of $N_{lg} \propto h^{1,1}$ can be explained from the weight system: As $n_\xi = \sum_{i}^4 n_i$ becomes large, a large number of lines has to be added to the weight matrix to make the threefold singularity free which generically increases the number of K\"ahler moduli. 
\begin{table}[t!]
\centering
  \begin{tabular}{l|ccccccc|c}
  $N_{lg}$ & 12-60 & 61-210 & 211-360 & 361-510 & 511-660 & 661-806 & 807-2330 & $\sum$\\
  \hline
  $\#_{\text{gen.}}$ & 1964 & 4101 & 1435 & 592 & 313 & 162 & 246 & 8813\\
  $\#_{\text{cons.}}$ & 889 & 1590 & 409 & 115 & 30 & 7 & 0 & 3040\\
  \end{tabular}
  \caption{The distribution of $N_{lg}$. The first line denotes the binning, while the second and third line contain the number of models in the respective $N_{lg}$ intervals for the general and conservative set, respectively.}
  \label{Xin1vsh11_tab}
\end{table}

We conclude this section with the remark that the possibility to engineer large enough gauge groups to obtain a large volume in the framework of K\"ahler uplifting is a generic feature of the landscape region we have analyzed.

\section{The type IIB perspective of $\mathbb{CP}_{11169}^4[18]$} \label{IIBperspectiveEx}

In this section, we present an explicit example of a brane and gauge flux setup on a threefold of the landscape region studied in section~\ref{lggr_sec}. To keep the analysis tractable we study a threefold with small $h^{1,1}$. Looking at figure~\ref{Xin1vsh11}, we see that there is a model with $h^{1,1} = 2$ that has $N_{lg} = 27$, which is close to the critical value $N_{lg}^{\text{crit.}} = 30$. This is the Calabi-Yau threefold $X_3$ that is a degree 18 hypersurface in $\mathbb{CP}^4_{11169}$ (it is usually denoted as $\Pex$). The corresponding weight system of the ambient toric space after resolving the singularities is
\begin{equation}
X_4^{\text{amb}} :\quad
\begin{array}{cccccc}
u_1 & u_2 & u_3 & u_4 & u_5 & \xi\\ 
\hline 
1 & 1 & 1 & 6 & 0 & 9\\
0 & 0 & 0 & 1 & 1 & 2\\
\end{array}\quad .
\label{WSPex}
\end{equation}
The two lines determine the two  scaling equivalence relations that the coordinates satisfy (see eq.~\eqref{equivrelamb}).
The Stanley-Reisner (SR) ideal encodes which homogeneous coordinates are not allowed to vanish simultaneously in the toric variety. For the four dimensional ambient space eq.~\eqref{WSPex} it is given by
\begin{equation}
 SR_{X_4^{\text{amb}}}=\{u_1 u_2 u_3, u_4 u_5 \xi\}\,.
\end{equation}

We will see in section~\ref{dS_sec} that on this  manifold we can stabilize the two K\"ahler moduli to values corresponding to a volume $\Vol \simeq 52$. In constructing an explicit brane and gauge flux setup on $X_3$ we address the following issues that are crucial in constructing a global model \cite{Blumenhagen:2007sm,Blumenhagen:2008zz,Collinucci:2008sq,Cicoli:2011qg,Cicoli:2012vw}:
\begin{itemize}
\item The choice of the orientifold involution determines the class of the O7-plane. D7-tadpole cancellation then implies $[D7]=-8[O7]$, fixing the degrees of the polynomial defining the D7-brane configuration. Requiring the presence of a D7-brane stack on $D_1$ with maximal gauge group rank $N_{lg}$ might force the defining polynomial to factorize further, leading the presence of another large rank stack (section~\ref{secO7D7}). Due to the swiss cheese structure of the volume form, this 
might destroy the large volume approximation and one has to check that this does not happens.
\item To lift unwanted zero modes that might destroy gaugino condensation on some D7-brane stacks, one needs to `rigidify' the wrapped non-rigid toric divisors (in the present case $D_1$). To do this, the gauge flux has to be properly adjusted (section~\ref{D1rigid_sec}).
\item To avoid the introduction of additional zero modes, due to non-zero gauge flux of the pull-back type (possibly forced by Freed-Witten anomaly cancellation~\cite{Minasian:1997mm,Freed:1999vc}), 
one has to choose such a flux in a proper way (section~\ref{sec_D1D5int}).
\item One has to saturate the D3-tadpole cancellation condition (section~\ref{D3tadpoleIIB_sec}).
\end{itemize}
In this specific model, we do not have to worry about the D5-tadpole cancellation. In fact, we will choose an orientifold involution with no odd $(1,1)$-forms, i.e. $h^{1,1}_-=0$. Hence, the D5-charge induced by gauge fluxes on D7-branes is automatically cancelled (there are no other source of D5-charge).
In the present model, also K-theoretic torsion charges are cancelled~\cite{Witten:1998cd,Moore:1999gb}.~\footnote{According to the probe argument presented in~\cite{Uranga:2000xp}, such charges are cancelled in a given set-up if any probe invariant D7-brane with gauge group $SU(2)$ has an even number of zero modes in the fundamental representation. This is realized in our model, as we always have an even number of D7-branes in the chosen stacks.}

\subsection{Geometric set-up and orientifold involution}

Before we go through the points listed above one by one, let us mention some geometric properties of $X_3$ that will be needed during the following analysis.  The Calabi-Yau is a hypersurface in the ambient space \eqref{WSPex}, defined by the equation
\begin{equation}\label{X3eq}
 \xi^2 = P_{18,4}(u_i)  \equiv u_5\, Q_{18,3}\,,  
\end{equation}
where $Q_{18,3} \equiv u_5^3 P_{18} + u_5^2 u_4 P_{12} + u_5 u_4^2 P_6 +  u_4^3$. The factorization of the polynomial $P_{18,4}$ is enforced by its weights.
The Hodge numbers of $X_3$ are $h^{1,1}= 2$ and $h^{2,1} = 272$. The (holomorphic) orientifold involution is given by
\begin{equation}\label{invX3}
 \xi \mapsto -\xi\:.
\end{equation}
This involution has $h^{1,1}_-=0$ and then the number of invariant K\"ahler moduli is $h^{1,1}_+=h^{1,1}=2$ (as one can find a two dimensional basis of $H^{1,1}$ in which each divisor is invariant under the orientifold involution). The orientifold plane is located at the fixed point locus of the involution \eqref{invX3}. In our case, the codimension three fixed locus is empty and hence we do not have O3-planes. On the other hand we have O7-planes on the codimension one fixed locus at $\xi=0$. Looking at equation \eqref{X3eq}, we see that this locus splits into two pieces. The corresponding four-cycles are given by the following equations in the ambient fourfold
\begin{equation}\label{O7planes}
 O7_{u_5}\,: \,\, \{\xi=0 \, \cap \, u_5=0\} \,, \qquad\qquad O7_{Q}\,: \,\, \{\xi=0 \, \cap \, Q_{18,3}=0\}\:.
\end{equation}
These hypersurfaces in $X_3$ are not complete intersections of one equation with the defining CY equation \eqref{X3eq}, but are four-cycles of the ambient fourfold that intersect non-transversely the CY $X_3$ and then are four-cycles also in $X_3$. Using the SR-ideal, one sees that these two four-cycles do not intersect each other on the Calabi-Yau $X_3$.
One can show that the divisor $O7_{u_5}$ is a rigid divisor in the threefold $X_3$.
Its homology class in the ambient space  is $D_\xi\cdot D_5=[X_3]\cdot \frac{D_5}{2}$. We see that the class of this integral four-cycle in the threefold is $D^{\rm fix}_5=\frac{D_5}{2}$. So we can use it as an element of an integral basis.

The threefold described so far  is the double cover of the three dimensional base manifold
\begin{equation}
B_3 :\quad
\begin{array}{ccccc}
u_1 & u_2 & u_3 & u_4 & u_5\\ 
\hline 
1 & 1 & 1 & 6 & 0\\
0 & 0 & 0 & 1 & 1\\
\end{array}\quad .
\label{WSBex}
\end{equation}
The toric divisors of $B_3$, defined by the equations $u_i=0$ will be called $\hat{D}_i$, in order to distinguish them from their double covers  $D_i$ in $X_3$ (given by the complete intersection $\{u_i=0\}\cap \{\xi^2-P_{18,4}=0\}$ in $X_4^{\rm amb}$).

The first Chern class of the base $B_3$ is $c_1(B_3)=\Kbar = 9 [\hat{D}_1] + 2 [\hat{D}_5]$ and its SR ideal is given by
\begin{equation}
 SR_{B_3}=\{u_1 u_2 u_3, u_4 u_5\}\,.
\label{SRB}
\end{equation}
Eq.~\eqref{SRB} can be used to derive triple intersections of the base divisors~\cite{Denef:2008wq,Denef:2004dm}
\begin{equation}
 \hat{D}_1^3 = 0\,, \qquad \hat{D}_1^2 \hat{D}_5 = 1\,, \qquad \hat{D}_1 \hat{D}_5^2 = -6\,, \qquad \hat{D}_5^3 = 36\,,
\label{baseints}
\end{equation}
where we have chosen the Poincar\'e duals of $\hat{D}_1$ and $\hat{D}_5$ as a basis of $H^{1,1}(B_3,\mathbb{Z})$. If the divisors intersect away from the fixed point locus, 
the intersections in the double cover threefold $X_3$ are simply twice the intersections in the base. 
Before we have seen that their double covers $D_1,D_5$ do not form an integral basis of $H^{1,1}(X_3,\mathbb{Z})$, i.e.\ there are integral divisors of $X_3$ that are rational combination of $D_1$ and $D_5$. Instead, an integral basis is given by $D_1,D^{\rm fix}_5$. Keeping in mind that $X_3$ is a double cover of $B_3$, that $D_1,D_5$ are the double covers of $\hat{D}_1,\hat{D}_5$ and that $D_5^{\rm fix}=\frac12 D_5$, we can determine the intersection numbers on $X_3$:
\begin{equation}
 D_1^3 = 0\,, \qquad D_1^2 D_5^{\rm fix} = 1\,, \qquad D_1 {D_5^{\rm fix}} ^2 = -3\,, \qquad {D_5^{\rm fix}} ^3 = 9\,.
\label{3foldints}
\end{equation}

One can check these results by realizing that the threefold $X_3$ can also be described as a hypersurface in the toric ambient space  
\begin{equation}
X^{\text{amb},123}_{4} :\quad
\begin{array}{cccccc}
u_1 & u_2 & u_3 & x & y & z\\ 
\hline 
1 & 1 & 1 & 6 & 9 & 0\\
0 & 0 & 0 & 2 & 3 & 1\\
\end{array}\quad \:,
\label{WSPex123}
\end{equation}
with Stanley-Reisner (SR) ideal 
\begin{equation}
 SR_{X^{\text{amb},123}_{4}}=\{u_1 u_2 u_3, x\,y\,z \}\,.
\end{equation}
We have used \textit{PALP} to verify that the Calabi-Yau threefolds described by the respective hypersurfaces in the ambient toric varieties~\eqref{WSPex} and~\eqref{WSPex123} are equivalent. At the level of their defining polytopes one sees this by finding that the normal forms~\cite{Kreuzer:1998vb} of the two polytopes are identical.\footnote{We thank A.~P.~Braun for discussion on this point.}

The defining equation of $X_3$ as an hypersurface in $X_4^{{\rm amb}123}$ is 
\begin{equation}\label{X3eq123}
 y^2 = x^3 + f_{12}(u_1,u_2,u_3) x \,z^4 + g_{16}(u_1,u_2,u_3) z^6 \:.
\end{equation}
The orientifold involution $\xi\mapsto -\xi$ is mapped to $y\mapsto
-y$. The fixed locus in the ambient space is given by two components,
i.e.\ $y=0$ and $z=0$ (the last one can be found once we apply the
involution plus the second line equivalence relation in
\eqref{WSPex123}). These loci intersect
transversely the Calabi-Yau $X_3$ and the intersections are connected. So we again find that the fixed point set in $X_3$ is given by two disconnected components. We can identify $D^{\rm fix}_5=D_z$ and $D^{\rm fix}_Q=D_y$. Furthermore, the divisors $D_{i=1,2,3}$ are identified with the corresponding ones in $X_4^{\text{amb}}$ \eqref{WSPex}. By using \textit{PALP}, and making the given identifications, we again obtain the triple intersections \eqref{3foldints}. Moreover we obtain the second Chern class of the threefold:
\begin{equation}
 c_2(X) = 102 D_1^2 + 69 D_1 D_5^{\rm fix} + 11 {D_5^{\rm fix}}^2\,.
\end{equation}

Expanding the K\"ahler form of $X_3$ as $J= t_1 D_1 + t_5 D_5^{\rm fix}$, we find an approximately swiss cheese volume of the threefold
\begin{equation}
 \Vol = \frac{1}{6} \int_X J \wedge J \wedge J = \frac{1}{2} \left( t_1^2 t_5 - 3 t_1 t_5^2 + 3 t_5^3 \right) = \sqrt{\frac{2}{3}} \left(\Vol_1 + \frac{1}{3}\Vol_5 \right)^{3/2} - \frac{\sqrt{2}}{9} \Vol_5^{3/2}\,,\label{Volex}
\end{equation}
with $\Vol_i = \partial \Vol / \partial t_i$. The K\"ahler cone is given by 
\begin{equation}
t_1-3\,t_5>0\,, \qquad t_5>0   \:.
\end{equation}

Finally, we list the hodge numbers, arithmetic genus $\chi_0$ and Euler number $\chi$ of divisors $D_1$ and $D_5^{\rm fix}$ in the threefold in Table~\ref{chi0D1D5IIB}. The results have been obtained by means of \textit{PALP} and \textit{cohomCalg}~\cite{Blumenhagen:2010pv,cohomCalg:Implementation}.
\begin{table}[ht!]
\centering
  \begin{tabular}{c|cccccc}
   & $h^{0,0}$ & $h^{1,0}$ & $h^{2,0}$ & $h^{1,1}$ & $\chi_0$ & $\chi$\\
  \hline
  $D_1$ & 1 & 0 & 2 & 30 & 3 & 36\\
  $D_5^{\rm fix}$ & 1 & 0 & 0 & 1 & 1 & 3\\
  \end{tabular}
  \caption{Hodge numbers, arithmetic genus $\chi_0$ and Euler number $\chi$ of $D_1$ and $D_5$.}
  \label{chi0D1D5IIB}
\end{table}
$D_5^{\rm fix}$ is rigid and hence fulfills the sufficient condition to contribute to the gaugino condensation superpotential. $D_1$ is not rigid, but one can choose a proper gauge flux on the wrapped D7-branes, that fixes the 
$h^{2,0}$ deformations.
We will perform this calculation in section~\ref{D1rigid_sec}.

\subsection{D7-brane configuration} \label{secO7D7}

Now, we discuss the inclusion of D7-branes, following the general procedure discussed in Section~\ref{D7IIBgen_sec}. To cancel the D7-charge of the O7-planes at $\xi=0$, the equation describing the D7-brane configuration is given by (see eq.~\eqref{WhitneyBr})
\begin{equation}
 \eta_{36,8}^2-\xi^2 \chi_{54,12} = 0 \,, 
\label{onewhitneyX}
\end{equation}
where the degrees of the $\eta$ and $\chi$ polynomials are dictated by the degrees of $\xi$ and by the requirement that $[D7]=-8[O7]=-8D_\xi$. To realize an $Sp(N_1)$ gauge group on the invariant divisor $D_1$ with $N_1 = N_{lg} = 27$ one takes
\begin{equation}\label{etachiSp24}
 \eta_{36,8} = u_i^{27} \tilde \eta_{9,8}\,,\qquad \qquad \quad \chi_{54,12} = u_i^{54} \tilde \chi_{0,12}\,.
\end{equation}
Eq.~\eqref{onewhitneyX} then becomes
\begin{equation}
 u_1^{54} \left(\tilde \eta^2_{9,8} - \xi^2 \tilde \chi_{0,12} \right) = 0\,.
\end{equation}
Since $\tilde \chi_{0,12} = u_5^{12}$, the Whitney brane splits into a brane and image brane due to the factorization $(\tilde \eta_{9,8} - \xi u_5^6)(\tilde \eta_{9,8} + \xi u_5^6)$. The Whitney brane can only carry a flux that is `trivial' on the Calabi-Yau threefold (i.e.\ its Poincar\'e dual non-trivial two-cycle on the D7-brane world volume is homologically trivial on the Calabi-Yau threefold), contrary to its split branes that in general carries a flux necessary for Freed-Witten anomaly cancellation. However, a non-trivial flux would generate additional (chiral) zero modes. We avoid them by choosing a degree of the Whitney brane polynomial, such that it does not split into a brane and its image. Furthermore, for the calculation of D3 charge of the Whitney brane, see Section~\ref{D3tadpoleIIB_sec}, we sacrifice two more gauge group ranks on $D_1$, realizing an $Sp(24)$ gauge group:
\begin{equation}
 u_1^{48} \left(\tilde \eta^2_{12,8} - \xi^2 \tilde \chi_{6,12} \right) = 0\:.
\end{equation}

We note that the polynomial $\tilde \eta_{12,8}$ is forced to factorize as $\tilde \eta_{12,8} = u_5^8\, \tilde \eta_{12} + u_5^7 u_4\, \tilde \eta_6 + u_5^6 u_4^2 = u_5^6\, \tilde \eta_{12,2}$, while the polynomial $\tilde \chi_{6,12}$ must have the form $\tilde \chi_{6,12} = u_5^{12}\, \tilde \chi_6 + u_5^{11} u_4 \tilde\chi_0$. In the following we will tune the parameter $\tilde \chi_0$ to zero. In this way the D7-brane configuration is described by the equation
\begin{equation}
 u_1^{48} u_5^{12} \left( \tilde \eta^2_{12,2} - \xi^2\, \tilde \chi_{6,0} \right) = 0\:.
\end{equation}

Hence, we see that requiring the factorization of the D7-brane
equation in order to produce an $Sp(24)$ stack enforces a further
factorization, in this case of $u_5$. Since $D_5=2D_5^{\rm fix}$ lies
on the O7-plane, this generates an $SO(24)$ gauge group.\footnote{If
  we do not take $\chi_0=0$, one has a factorization as $u_1^{48}
  u_5^{12} \left( \tilde \eta^2_{12,2} - Q_{18,3}\, \tilde \chi_{6,1}
  \right) = 0$, where we used $\tilde \chi_{6,12} = u_5^{12}\, \tilde
  \chi_6 + u_5^{11} u_4= u_5^{11}\, \tilde \chi_{6,1}$ and $\xi^2 =
  u_5 Q_{18,3}$. 
From this we would conclude that there is an $SO(24)$ gauge group on
$D_5^{\rm fix}$. However, as we will see in the F-theory language, the actual gauge group is $SO(23)$.}

The same conclusions on the factorization can be obtained using the description of $X_3$ as a hypersurface in $X_{4}^{{\rm amb},123}$ \eqref{WSPex123}. Remember that in that description the orientifold locus is given by $D_y+D_z$. To cancel the D7-tadpole, we need a D7-brane configuration wrapping the divisor class $8(D_y+D_z)=8(9D_1+4D_z)$, i.e.\ it is given by the equation $P_{72,32}=0$. A  polynomial with these degrees must factorize as $P_{72,32}=z^8P_{72,24}$. Hence there are eight branes wrapping the orientifold-plane divisor $D_z$ (realizing a group $SO(8)$ if one does not require further factorizations). The D7-tadpole generated by the O7-plane wrapping $D_y$ is cancelled by a Whitney brane, defined by the equation $\eta_{36,12}^2-y^2\chi_{54,18}=0$. As above, we require an $Sp(24)$ stack on $u_1=0$: $\eta_{36,12}=u_1^{24}\eta_{12,12}$ and $\chi_{54,18}=u^{48}\chi_{6,18}$. The polynomial $\eta_{12,12}$ and $\chi_{6,18}$ must factorize as $\eta_{12,12}=z^8\hat{\eta}_{12,4}$ and $\chi_{6,18}=z^{16}\hat{
\chi}_{6,2}$ (as above, we will take $\hat{\chi}_{6,2}=z^2\hat{\chi}_{6,0}$). We see that we have 16 more D7-branes that wrap the O7-plane divisor $D_z$, realizing an $SO(16+8)=SO(24)$ gauge group.

\subsection{Rigidifying $D_1$ by gauge flux} \label{D1rigid_sec}

In this section we construct explicitly a gauge flux on the $Sp(24)$ stack wrapping $D_1$, 
that fixes all the deformation moduli of these branes. The equation describing $D_1$ is $u_1=0$, which can be
deformed to $u_1+\zeta_2 u_2 +\zeta_3 u_3=0$. We see that we have two deformation moduli, consistent with the
fact that $h^{2,0}(D_1)=2$. We need to lift such zero modes in order to avoid destroying gaugino condensation on the D7-brane stack wrapping $D_1$.

The rigidifying flux will be taken such that it is not a pull-back of a $CY_3$ two-form. 
In this way, it will not generate additional chiral matter and will not enter
in the D-term constraints. On the other hand, since $h^{2,0}(D_1) \not =0$, 
this flux will constrain the holomorphic embedding of the D-brane, 
by the F-term constraint ${\cal F}^{2,0}=0$. 
This type of gauge flux  was introduced first in \cite{Bianchi:2011qh} to fix the deformation moduli of a non-rigid divisor wrapped by an E3-instanton. The use of such a flux to fix the deformation moduli of D7-branes such that they can support gaugino condensation was suggested in \cite{Cicoli:2012vw}. Here we make this flux explicit, proving that it indeed fixes the unwanted deformations and computing its D3-tadpole contribution. 

To construct the flux, we follow the procedure described in \cite{Bianchi:2011qh}: 
We have to identify  holomorphically embedded curves in the CY threefold that do not admit holomorphic deformation.
In fact, the condition ${\cal F}^{2,0}=0$ means that the Poincar\'e dual two-cycle 
in $D_1$ remains holomorphic when the divisor is deformed in 
the threefold. If we cannot deform the curves, then some of the deformation moduli of $D_1$ are fixed.

We will work in the case in which the threefold is described by a hypersurface in the ambient space \eqref{WSPex123} with the defining equation \eqref{X3eq123}. 
The rigid curves cannot be described by the intersection of one equation with \eqref{X3eq123} and $u_1=0$. In fact these will be always holomorphic as we deform the $D_1$ equation. To visualize these rigid holomorphic curves in an algebraic way, we have to parametrize $g_{16}$ appropriately, i.e.\ $g_{16}=\psi_8^2 + u_1 \tau_{15}$ where $\psi_8$ and $\tau_{15}$ are polynomials of degree 8 and 15 in the $u_i$ coordinates.\footnote{An analogous ansatz was done in \cite{Braun:2011zm} for an elliptically fibered Calabi-Yau fourfold, in order to find rigid four-cycles inside the fourfold.} In this way we can write the CY equation as
\begin{equation}
 (y - \psi_8) (y + \psi_8) = x (x^2 + f_{12}\, z^4) + u_1 \tau_{15}\,z^6 \:.
\end{equation}
The rigid curve we are interested is the $\mathbb{P}^1$
\begin{equation}\label{rigidC}
 {\cal C} : \qquad y = \psi_8 \qquad \cap \qquad x=0 \qquad \cap
 \qquad u_1=0\ .
\end{equation}
Through exact sequences (following \cite{Katz}), 
one shows that its normal bundle in the threefold is ${\cal O}(-1)\oplus {\cal O}(-1)$ and then it has no holomorphic sections, proving that this curve is rigid in the threefold.

Using the curve \eqref{rigidC}, we now construct the gauge flux:
\begin{equation}\label{RigFlux}
 {\cal F}_{D_1} = \tfrac12 D_1 - B + {\cal C} - s = {\cal C} - s \:,
\end{equation}
where we will choose the B-field $B$ such that $B-\frac12 D_1=0$ to cancel the Freed-Witten flux $\frac{D_1}{2}$.
With abuse of notation, in eq.~\eqref{RigFlux} we call ${\cal C}$ the two-form Poincar\'e dual to the curve~\eqref{rigidC} in $D_1$.
$s$ is the pull-back of a threefold two-form, i.e.\ its dual cycle is the intersection of one equation $P_s = 0$
with \eqref{X3eq123} and $u_1=0$. Deforming the $D_1$ equation, $P_s=0$ is not modified.
We will choose $s$ such that the two-cycle dual to this flux is trivial in 
the CY threefold. This corresponds to the flux being orthogonal to the two-forms that are pulled back from the CY threefold. This is required
in order to prevent chiral matter generated by the flux. $s$ is by construction of type $(1,1)$. So the condition $ {\cal F}^{2,0}_{D_1}=0$ is
equivalent to ${\cal C}^{2,0}=0$. Requiring that ${\cal C}$ remains of type $(1,1)$ as $D_1$ is deformed, is the same as requiring that
the curve \eqref{rigidC} is contained in the deformed divisor $u_1+\zeta_2 u_2 +\zeta_3 u_3=0$. But we see that this happens 
only if $\zeta_2=\zeta_3=0$, i.e.\ the deformation moduli are fixed by this flux.

Let us determine the homology class of $s$ in the CY.\footnote{In the case we are considering, two cycles of the Calabi-Yau threefold are homologous if their push-forward in the ambient fourfold are homologous.} We want that $[s]=[{\cal C}]|_{CY}$, such that ${\cal F}_{D_1}$ is trivial in the CY.
The class of ${\cal C}$ in the ambient space is 
\begin{equation}
[{\cal C}]=D_y \cdot D_x \cdot D_1 = \frac12 [CY] \cdot D_x \cdot D_1 = 
[CY] \cdot (3D_1+D_z)\cdot D_1 \:. 
\end{equation}
So $[s] = [{\cal C}]|_{CY} = (3D_1+D_z)\cdot D_1$.

Now we compute the self-intersection ${\cal F}_{D_1}^2$, that enters into the D3-charge of the flux
\begin{equation}
-\frac12 \int_{D_1}{\cal F}_{D_1}^2 = -\frac12 \int_{D_1} ({\cal C} - s)\cdot ({\cal C})= -\frac12\int_{D_1} ({\cal C}^2 - s^2)\:,
\end{equation}
where we have used the fact that ${\cal F}_{D_1}$ is orthogonal to every pulled back two-form (and then also to $s$).
The only difficult part to compute is ${\cal C}^2$. To do this we apply the following relation:
\begin{equation}
 {\cal C} \cdot {\cal C}|_{D_1} = \int_{\cal C} {\cal C} = \int_{\cal C} c_1(N|_{{\cal C}\subset D_1})\:,
\end{equation}
where we have used that the normal bundle of ${\cal C}$ in $D_1$ is a line bundle whose first Chern class 
is given by the class of the curve ${\cal C}$ in $D_1$. The class $c_1(N|_{{\cal C}\subset D_1})$ can be computed
via the following exact sequence:
\begin{equation}
 0 \rightarrow N|_{{\cal C}\subset D_1} \rightarrow N|_{{\cal C}\subset Y_4} \rightarrow N|_{D_1\subset Y_4} \rightarrow 0 \:.
\end{equation}
From this we get that $c_1(N|_{{\cal C}\subset D_1})=c_1(N|_{{\cal C}\subset Y_4})-c_1(N|_{D_1\subset Y_4})$.
The classes on the right hand side can be easily determined by the fact that the objects involved are complete intersections in 
a toric space:
\begin{eqnarray}
 c_1(N|_{{\cal C}\subset D_1})&=& c_1(N|_{{\cal C}\subset Y_4})-c_1(N|_{D_1\subset Y_4})\nonumber\\
     &=&    (D_y + D_x + D_1) - (2D_y + D_1) = D_x-D_y\\
     &=&    -3D_1-D_z \:.\nonumber
\end{eqnarray}
Then
\begin{align}
\begin{aligned}
 {\cal C} \cdot {\cal C} |_{D_1}&= \int_{\cal C} c_1(N|_{{\cal C}\subset D_1}) = \int_{Y_4} [{\cal C}] \cdot c_1(N|_{{\cal C}\subset D_1})\,=\,
   \int_{Y_4} D_y\cdot D_x\cdot D_1 \cdot (-3D_1-D_z) \\
    &= \int_{X_3} \tfrac12 (6D_1+2D_z) \cdot D_1 \cdot (-3D_1-D_z) \,=\,
   - \int_{X_3}D_1\cdot (6D_1\cdot D_z + D_z^2) = -3 
\end{aligned}
\end{align}
(Using analogous techniques, one can also compute $c_1({\cal C})$ and prove that ${\cal C}$ is indeed a $\mathbb{P}^1$.)

More easily, we can compute 
\begin{equation}
\int_{D_1}s^2 = \int_{X_3}D_1\cdot (3D_1+D_z)^2 = 3\:.  
\end{equation}
We are ready to compute the D3-charge induced by this flux:
\begin{equation}
-\frac12 \int_{D_1}{\cal F}_{D_1}^2 =  -\frac12\int_{D_1} ({\cal C}^2 - s^2) = 3 \:.
\label{intF1F1D1}
\end{equation}
We will switch on this flux along all the 24 branes making up the $Sp(24)$ stack, and its image along the 24 image branes. It is
easy to prove that, as expected,  ${\cal F}'_{D_1}\equiv {\cal C}'-s' = - ({\cal C}-s) = -{\cal F}_{D_1}$, where ${\cal C}'$ is given by the equations
in \eqref{rigidC} with the substitution $\xi\mapsto -\xi$ and the class of $s'$ is such that ${\cal F}'_{D_1}$ is trivial in the CY $X_3$.
This (diagonal) flux breaks the $Sp(24)$ gauge group to $U(24)$ (the $U(1)$ factor remains massless, 
because of the triviality of the flux in $X_3$).

In conclusion, the D3-charge of the gauge flux needed to rigidify all the D7-branes in the $Sp(24)$ stack is
\begin{equation}
 Q_{D3}^{{\cal F}_{D_1}}= - \frac{24}{2}\int_{D_1}{\cal F}_{D_1}^2 - \frac{24}{2}\int_{D_1}{\cal F'}_{D_1}^2 = - 2\times \frac{24}{2}\int_{D_1}{\cal F}_{D_1}^2 = 144\:.\label{D3gaugeflux}
\end{equation}

\subsection{Avoiding D-terms and zero-modes from matter fields} \label{sec_D1D5int}

In Section~\ref{secO7D7}, we introduced D7-brane stacks on the divisors $D_1$ and $D_5$. If the branes carry non-zero flux, we have to
worry about possible charged matter fields arising at the intersection of the two D7-brane stacks or from the D7-brane bulk spectrum. These zero modes might force the contribution to the superpotential from gaugino condensation to be zero.
These problematic fluxes also generate K\"ahler moduli dependent Fayet-Iliopoulos (FI) terms $\xi_i$ \cite{Dine:1987xk,Jockers:2005zy,Haack:2006cy}. This would introduce a D-term potential for the K\"ahler moduli. However, the method of K\"ahler uplifting which we use requires a pure F-term potential.

In the following, we show that additional zero-modes and D-terms can be avoided for an appropriate choice of gauge flux $F$ on the branes wrapping the divisors $D_1$ and $D_5^{\rm fix}$. The gauge flux $F$ combines with the pull-back of the bulk B-field on the wrapped four-cycle to give the gauge invariant field strength
\begin{equation}
\mathcal{F} = F - B\,. 
\label{ginvfl}
\end{equation}
The number of additional zero modes and the K\"ahler moduli dependent Fayet-Iliopoulos terms appearing in D-terms are given by integrals of the form
\begin{equation}
 \int_{D_i} \mathcal{F}_{D_i} \wedge D = \int_{X_3} D\wedge  D_i \wedge \mathcal{F}_{D_i} \,,
\label{DiFDi}
\end{equation}
where $D$ is an arbitrary divisor in the threefold $X_3$. If it is possible to choose the fluxes $\mathcal{F}_{D_1}$ and $\mathcal{F}_{D_5^{\rm fix}}$ such that eq.~\eqref{DiFDi} vanishes for $i=1,5$ these fluxes do not have any problematic consequences.
In particular, an integral such as \eqref{DiFDi} vanishes if the flux $\mathcal{F} _{D_i}$ is orthogonal to the two-forms of $D_i$ that are pull-backed from the Calabi-Yau threefold $X_3$.

When turning on gauge flux one has to make sure that the Freed-Witten anomaly~\cite{Minasian:1997mm,Freed:1999vc} is canceled, i.e.\ the gauge flux on a brane wrapping divisor $D$ has to satisfy
\begin{equation}
 F+\frac{c_1(D)}{2} \in H^2(X_3,\mathbb{Z})\,.
\end{equation}
If the divisor $D$ is non-spin, its first Chern class $c_1(D)$ is odd and $F$ cannot be set to zero.
On the other hand, the expression appearing in the physical quantities is the gauge invariant flux eq.~\eqref{ginvfl}. By choosing the B-field appropriately, one can make this invariant flux equal to zero. 
For a set of D7-brane stacks wrapping non-intersecting divisors, the global B-field can be chosen such that the pull-back on all such divisors make ${\cal F}=0$ for all the stacks. However, for intersecting stacks this is not possible in general.  We now prove that our case is not generic in this respect: We can choose the B-field such that both ${\cal F}_{D_1}=0$ and  ${\cal F}_{D_5^{\rm fix}}=0$.

The cancellation of the Freed-Witten anomaly is always satisfied if the flux that is switched on is given by
\begin{equation}
 F = f_1\, D_1 + f_5\, D_5^{\rm fix} + \frac{D}{2}\qquad \text{with} \qquad f_1,f_5 \in \mathbb{Z}\,,
\label{FreedWitten}
\end{equation}
where we have used that $D_1$ and $D_5^{\rm fix}$ form a basis of $H^2(X_3,\mathbb{Z})$ and that $c_1(D) = - D$ in a Calabi-Yau threefold. We will only turn on diagonal fluxes on the stacks wrapping $D_1$ and $D_5^{\rm fix}$, i.e.\ the flux on each brane of a stack is the same (and the opposite on all the image branes). If we took non-diagonal fluxes into account this should be reflected by an additional index, for example $F^\alpha$ where $\alpha$ enumerates the branes.  

As mentioned above we can use the B-field to fix one of the gauge fluxes $\mathcal{F}_{D_1}$ or $\mathcal{F}_{D_5^{\rm fix}}$ to zero. In particular the Freed-Witten gauge flux, i.e.\ $\frac{D_i}{2}$, cannot be cancelled by an integral shift and then a half-integrally quantized $B$-field is needed. We choose $B=\frac{D_1}{2}$ (up to an integral two-form), so that choosing $F_{D_1}=\frac{D_1}{2}$ (up to the same integral two-form) one trivially realizes ${\cal F}_{D_1}=0$. 
The chosen B-field leads the gauge invariant flux on $D_5^{\rm fix}$ to be $\mathcal{F}_{D_5^{\rm fix}} = F_{D_5^{\rm fix}} - \frac{D_1}{2}$. We will now show that $F_{D_5^{\rm fix}}$ can be tuned such that the vanishing of eq.~\eqref{DiFDi} can also be accomplished for $D_i=D_5^{\rm fix}$. In particular, we will show that the pull-back of $\frac{D_5^{\rm fix}}{2}-\frac{D_1}{2}$ on $D_5$ is zero.\footnote{ 
In principle, one could also try to choose the B-field such that $\mathcal{F}_{D_5^{\rm fix}}=0$ (i.e. $B=\frac{D_5^{\rm fix}}{2}$), instead of $\mathcal{F}_{D_1}=0$. The problem with this choice is that the pull-back of $\frac{D_1}{2}-\frac{D_5^{\rm fix}}{2}$ on $D_1$ is not zero.}
Let us see this in detail. After the given choice of the B-field, the gauge-invariant field strengths are given as
\begin{align}\begin{aligned}
 \mathcal{F}_{D_1}&=0\,,\\
\mathcal{F}_{D_5^{\rm fix}}&=  \left(a-\frac{1}{2}\right)D_1 + \left(b+\frac{1}{2}\right)D_5^{\rm fix}\quad \text{with} \quad a,b \in \mathbb{Z}\,.
\end{aligned}\end{align}
Now it is easy to see that for the choice of fluxes
\begin{equation}
 a=2+3b\, \qquad \Rightarrow \qquad \mathcal{F}_{D_5^{\rm fix}}=\frac{1}{2} (1+ 2 b) (3 D_1 + D_5^{\rm fix})\,,
\label{gaugefluxcond}
\end{equation}
the pull-back of $\mathcal{F}_{D_5^{\rm fix}}$ on $D_5^{\rm fix}$ becomes trivial since
\begin{equation}
 \int_{X_3} D_5^{\rm fix} \wedge \mathcal{F}_{D_5^{\rm fix}} \wedge D \propto \int_{X_3} D_5^{\rm fix} \wedge (3 D_1 + D_5^{\rm fix}) \wedge D = 0\,,
\label{FD5trivial}
\end{equation}
for an arbitrary divisor $D = k_1 D_1 + k_5 D_5^{\rm fix} \in H^2(X_3,\mathbb{Z})$. In the last equality in eq.~\eqref{FD5trivial} we have used the triple intersections of $X_3$, eq.~\eqref{3foldints}.
Hence, we have shown that additional zero-modes as well as D-terms can be avoided by tuning the gauge flux on the brane stack on $D_5$.

\subsection{D3 tadpole cancellation condition} \label{D3tadpoleIIB_sec}

The D3 tadpole has to cancel for consistency. The compactification ingredients that induce a D3 charge are the (fluxed) D7-branes, the O7-planes, the D3-branes, the O3-planes and the RR and NS field strengths $F_3$ and $H_3$. 
The RR and NS fluxes and the D3-branes have a positive contribution:
\begin{equation}
 Q_{D3}^{F_3,H_3} = \frac{1}{2} \int_{X_3} F_3 \wedge H_3 \,,\qquad Q_{D3}(N_{D3}\times D3) = N_{D3}\,.
\label{QD3fluxes}
\end{equation}

In our case we do not have O3-planes, while we have O7-planes. Each 
O7-plane contributes negatively by
\begin{equation}
 Q_{D3}^{O7} = -\frac{\chi(O7)}{6}\,.
\end{equation}
In our construction we have two O7-planes (see eq.~\eqref{O7planes}), whose D3-charge sum up to
\begin{equation}
 Q_{D3}^{O7s} = Q_{D3}^{O7_{u_5}}+Q_{D3}^{O7_Q} = -\frac{\chi(O7_{u_5})}{6}-\frac{\chi(O7_Q)}{6}= -\frac16 ( 3+549)=-92\,,
\label{D3O7pl}\end{equation}
where we used $\chi(O7_{u_5})=3$ and $\chi(O7_Q)=549$.

A stack of $N_i$ D7-branes and their $N_i$ images wrapping a divisor $D_i$ contributes to the total D3 charge positively via the gauge flux and negatively via a geometric contribution:
\begin{equation}
 Q_{D3}^{D7} (D_i) = 2 N_{i}\left( -\frac{1}{2} \int_{D_i} \mathcal{F}_{D_i} \wedge \mathcal{F}_{D_i} - \frac{\chi(D_i)}{24} \right)\,,
\end{equation}
where the overall factor two comes from sum over the stack and its image stack which have the same D3 charge.
For the brane-stacks  on $D_1$ and $D_5^{\rm fix}$ described in Section~\ref{secO7D7}, we obtain the following D3 tadpole:
\begin{align}
Q_{D3}^{\text{stacks}} =  &-\frac{2N_1}{2}\int_{D_1} (\mathcal{C}-s)^2 -2 N_{1} \frac{\chi(D_1)}{24} - 2 N_{5} \frac{\chi(D_5^{\rm fix})}{24} =  144-3 - 72 = 69 \,,\label{D3stacksO7}
\end{align}
where we have used $N_1 = 24$, $N_5 = 12$, eq.~\eqref{D3gaugeflux} for the D3-charge of $\mathcal{F}_{D_1}$ and the results of Table~\ref{chi0D1D5IIB}. 

The Whitney brane, defined by the equation $\eta^2-\xi^2\chi=0$, has a singular world volume. Thus we have to compute its contribution to the D3-tadpole indirectly. We use the fact that the Whitney brane wrapping a divisor class $D_W=2D_P$ can be seen as the recombination of a brane wrapping the invariant divisor $D_P$ with the image brane wrapping the same divisor class $D_P$. On this brane the flux is ${\cal F}_{D_P}=\frac12 D_P-S-B$, while on the image brane, it is given by $-{\cal F}_{D_P}$. Here $B$ is the $B$-field and $S$ an arbitrary integral class. After recombination only a flux that is trivial along the CY survives (see \cite{Collinucci:2008pf,Collinucci:2008sq}). 
In the recombination process, the RR-charges do not change. Hence, we
can compute the D3-charge in the more tractable brane/image-brane situation, as done in \cite{Collinucci:2008sq,Cicoli:2011qg}:
\begin{equation}\label{D3chWhitBr}
 Q_{D3}^W = -\int_{D_P} {\cal F}_{D_P}^2 - \frac{\chi(D_P)}{12}  \:.
\end{equation}

In order for the Whitney brane to be holomorphic (supersymmetric), $D_P$ must be such that $D_P-[O7]>0$ (i.e.\ it can be described by the
vanishing of a holomorphic equation).
The choice of $S$ in not completely arbitrary. It must satisfy the following constraints, in order for the Whitney brane to be stable against splitting:
\begin{equation}\label{ConstrWhitStab}
  \frac{[O7]}{2} \leq S+B \leq D_P- \frac{[O7]}{2} \:.
\end{equation}

We will compute the charge using the description of $X_3$ as a hypersurface in the ambient space $X_{4}^{{\rm amb},123}$ \eqref{WSPex123}.
The equation of the Whitney brane, after factorizing the $Sp(24)$ and the $SO(24)$ stacks is given by $\hat{\eta}^2_{12,4}-\xi^2\hat{\chi}_{6,2}$.
In our example the B-field is fixed by the vanishing of ${\cal F}_{D_1}$ to be $B=\frac{D_1}{2}$. The class of the Whitney brane is 
$D_W = 2D_P=2 (12 D_1+4 D_z)$, while $[O7]=D_y=9D_1+3D_z$. If we take $S=f_1 D_1+ f_z D_z$ ($f_1,f_z\in\mathbb{Z}$), the constraints \eqref{ConstrWhitStab} translate to $4\leq f_1\leq 7$ and $\frac32 \leq f_z \leq \frac52 $, i.e.\ $f_1=4,5,6,7$ and $f_z=2$.

Let us compute the Whitney brane D3-charge. The Euler number of $D_P$ is 
\begin{equation}
\chi(D_P)=\int_{X_3} D_P^3+D_P\cdot c_2(X_3)=984 \:,
\end{equation} 
while
\begin{equation}
\int_{D_P} {\cal F}_{D_P}^2=\int_{X_3}D_P (\frac{D_P}{2}-\frac{D_1}{2}-(f_1D_1+f_zD_z))^2=-(2f_1-11)^2 \:.
\end{equation}
Inserting these expressions in \eqref{D3chWhitBr} and using the possible values for $f_1$, we obtain two possible results for the charge of the Whitney brane, i.e.\ $Q_{D3}^W = -81$ and $Q_{D3}^W = -73$.

Taking into account the contribution from the O7-planes and the D7-brane stacks, eq.~\eqref{D3O7pl} and \eqref{D3stacksO7}, and the negative contribution from the Whitney brane, we obtain the following total D3-brane charge from our brane configuration:
\begin{equation}\label{totalD3charge}
 Q_{D3}^{\text{tot}} = Q_{D3}^{\text{O7s}} +Q_{D3}^{\text{stacks}} + Q_{D3}^W  = \begin{cases} -104 &\mbox{for } Q_{D3}^W = -81 \\ 
-96 & \mbox{for } Q_{D3}^W = -73\,. \end{cases}
\end{equation}

At this point, we have a fully consistent picture of the D brane and gauge flux setup in our threefold $X_3$ that ensures that gaugino condensation from the divisors $D_1$ and $D_z$ contributes to the superpotential of the four dimensional ${\cal N}=1$ effective supergravity. In particular, we have overcome the issues discussed in Section~\ref{lggr_sec}. Before we discuss moduli stabilization in a de Sitter vacuum in Section~\ref{dS_sec}, let us have a look at the brane and flux configuration from the F-theory point of view in the following Section~\ref{FthEx_sec}.

\section{The F-theory perspective of $\mathbb{CP}_{11169}^4[18]$} \label{FthEx_sec}

In this section, we revisit some results of the previous section by using the F-theory language.
We first discuss the D7-brane configuration in F-theory, according to Table~\ref{SpSUSO}. We consider the Calabi-Yau fourfold that is an elliptic fibration over the base manifold $B_3$ defined in eq.~\eqref{WSBex}.
We find that enforcing an $Sp(24)$ singularity on the divisor $\hat{D}_1$ of $B_3$ forces us to impose an $SO(24)$ singularity on $\hat{D}_5$. This agrees with the type IIB perspective discussed in Section~\ref{secO7D7}. 

Let us see this in detail. The Tate polynomials are sections of $\Kbar^i$ which we denote as $a_i = A_{9i,2i}$. By realizing an $Sp(24)$ gauge group on $D_1$ we need the following factorization of the Tate polynomials:
\begin{align}
\begin{aligned}
 &a_1 = A_{9,2} =u_5 A_{9,1}\,,\\ 
 &a_2 = A_{18,4} = u_5 A_{18,3}\,,\\
 &a_3 = u_1^{24} A_{3,6} =u_1^{24} u_5^6 A_{3,0} \,,\\ 
 &a_4 = u_1^{24} A_{12,8} = u_1^{24} u_5^6  A_{12,2}\,,\\ 
 &a_6 = u_1^{48} A_{6,12} = u_1^{48} u_5^{11} (u_4 A_{0,0}  + u_5 A_{6,0})\,,\label{aifacPex}
\end{aligned}
\end{align}
where on the RHS of~\eqref{aifacPex} we made the $u_5$ factorization explicit. Comparing with Table~\ref{SpSUSO}, we see that the $a_i$ in eq.~\eqref{aifacPex} are compatible both with an $SO(4\cdot 5+3=23)$ and an $SO(4\cdot5+4=24)$ singularity on $\hat{D}_5$. For generic polynomials $A_{p,q}$, the first is realized. The $SO(24)$ is present when the polynomial $A_{18,3} X^2 + A_{12,2} X + (u_4A_{0,0}+u_5 A_{6,0})$ factorizes modulo $u_5$ (i.e.\ modulo a polynomial that is divided by $u_5$). This happens if $A_{0,0}=0$, that is the choice we have done in the perturbative type IIB limit in Section~\ref{secO7D7} ($\chi=4a_6+a_3^2$).

Now we consider the fourfold where the $Sp(24)$ singularity is resolved along the lines described in Appendix~\ref{SpNgeom}. The resolution introduces a set of new divisors, the exceptional divisors $E_{2i-1}$ ($i=1,...,N$), that are $\mathbb{P}^1$ fibrations over the surface in the base $B_3$ where the fiber degenerates.
From the F-theory point of view, the gaugino condensation contribution to the superpotential is generated by M5-instantons wrapping the exceptional divisors that resolve the corresponding singularity~\cite{Katz:1996th}. In the presence of fluxes, the necessary condition for an M5-instanton wrapping a divisor $D$ to contribute to the superpotential is that $\chi_0(D)\geq 1 $, which is the known modification of the condition $\chi_0(D)=1$ without fluxes~\cite{Denef:2008wq,Kallosh:2005gs}. 

In Appendix ~\ref{SpNgeom}, we derive the following formula for the arithmetic genus of the exceptional divisors
\begin{align}
\begin{aligned}
 \chi_0(E_{2i-1}) &= \frac{1}{6} \int_{B_3} \hat{D} \wedge \left[c_2(B_3)+\Kbar \wedge \Kbar + 2 \hat{D} \wedge \hat{D} \right]= \chi_0(D);\,\, i=1,..,N-1\,,\\
 \chi_0(E_{2N-1}) &= \frac{1}{12} \int_{B_3} \hat{D} \wedge \left[c_2(B_3)+ (\Kbar - \hat{D}) \wedge (\Kbar - 2 \hat{D}) \right] = \chi_0(\hat{D})\,,
\end{aligned}\label{chi02}
\end{align}
where $\hat{D}$ is the divisor $\{p_{\hat{D}}=0\}$ on the base manifold $B_3$ where the singularity sits and $D$ is its double cover in $X_3$.

 In our example we imposed an $Sp(N_1=24)$ singularity on the divisor $\hat{D}_1$. Inserting
\begin{equation}
 \Kbar = 9 \hat{D}_1 + 2 \hat{D}_5 \,,\qquad\qquad
 c_2(B_3) = 21 \hat{D}_1^2 + 12 \hat{D}_1 \hat{D}_5 + \hat{D}_5^2\,,
\end{equation}
and the base intersections eq.~\eqref{baseints} into eq.s~\eqref{chi02} we find
\begin{equation}
 \chi_0(E_{2i-1}) = 3\,\,\text{for}\,\, i=1,\dots,N_1-1  \qquad \text{and} \qquad \chi_0(E_{2N_1-1}) = 1\,.
\end{equation}
We see that all of them satisfy the necessary condition for an M5-instanton to contribute to the superpotential in the presence of fluxes.
This agrees with what we found in type IIB language, where we have seen that this actually happens, i.e.\ switching on a proper gauge flux fixes the deformation on the wrapped divisor, leading to the possibility of having gaugino condensation.

Let us summarize the F-theory analysis. The D-brane configuration is equivalent to what we find in Section~\ref{secO7D7} in the IIB picture. Furthermore, we calculated the arithmetic genus of the exceptional divisors that correspond to the resolution of the $Sp(N)$ singularity for a general base manifold. For our example, the results show that one has to switch on gauge flux so that the divisor can carry a non-vanishing contribution from gaugino condensation to the superpotential.  The corresponding type IIB analysis was done in Section~\ref{D1rigid_sec}.

\section{Moduli stabilization in the large volume limit} \label{CSlargeVolume}

Let us define the four dimensional effective theory. In the absence of
$D$-terms the scalar potential is completely determined by the K\"ahler potential $K$ and the superpotential $W$ via
\begin{equation}
 V = e^{K} \left( K^{\alpha\bar{\beta}} D_\alpha W \overline{D_\beta W} - 3 |W|^2 \right)\,,\label{V4D}
\end{equation}
where $D_\alpha W = W_\alpha + K_\alpha W$, with subscripts $\alpha,\beta$ denoting the derivative w.r.t. complex scalar moduli fields $\phi_\alpha$. These are the K\"ahler moduli $T_i$, $i=1,\dots,h^{1,1}$, the dilaton $S$ and the complex structure moduli $U_a$, $a=1,\dots,h^{2,1}$.

The K\"ahler potential is a real function of these moduli \cite{Grimm:2004uq}:
\begin{equation}
K = - 2 \log \left( \Vol(T_i,\bar T_i) +  \frac{1}{2}\hat\xi(S,\bar S) \right) - \log (S+\bar{S}) - \log \left( - i \int \bar{\Omega}(\bar U_a) \wedge \Omega (U_a) \right) \,\,,\label{Kgen}
\end{equation}
where $\Vol$ denotes the volume and $\Omega$ the holomorphic three-form of the threefold while the dilaton dependent $\alpha'$ correction is given as
\begin{equation}
 \hat\xi(S,\bar S)= - \frac{\zeta(3)\,\chi}{4 \,\sqrt{2}\, (2 \pi)^3}   \, (S+\bar{S})^{3/2}\,,\label{xihatgen}
\end{equation}
with $\chi$ being the Euler number of the threefold and $\zeta(3) \simeq 1.202$.

The holomorphic superpotential
\begin{equation}
W = W_{0}(S,U_a) + \sum_i A_i(S,U_a)\, e^{- a_i T_i}\label{Wgen}\,,
\end{equation}
is the sum of a tree-level and a non-perturbative contribution. In the model we consider, the non-perturbative exponential term arises from gaugino condensation in pure non-Abelian 4D ${\cal N}=1$ super-Yang-Mills theories on stacks of D7-branes wrapping appropriate four-cycle divisors of the Calabi-Yau threefold. Here $A_i(S,U_a)$ denotes the one-loop determinant, and $a_i = 2\pi/N_i$ where $N_i$ is the Coxeter number of the corresponding gauge group singularity in F-theory. The perturbative term consists of the flux induced Gukov-Vafa-Witten superpotential~\cite{Gukov:1999ya}
\begin{equation}
  W_0(S,U_a) = \frac{1}{2 \pi} \int (F_3 - S H_3) \wedge \Omega(U_a) \,,
\end{equation}
with $S= s + i\, \sigma$, $s=1/g_s$ and $F_3$ and $H_3$ the RR and NS three-form field strength respectively.

As discussed in much recent work (see~\cite{Giddings:2001yu} for the original seminal result, and~\cite{Douglas:2006es,Grana:2005jc,Blumenhagen:2006ci} for recent review), turning on primitive imaginary self-dual (ISD) three-form fluxes generically leads to a supersymmetric stabilization of all complex structure moduli and the axio-dilaton at an isolated supersymmetric extremum in moduli space. The flux induced superpotential is independent of the K\"ahler moduli, which implies 
\beq
K^{i\bar\jmath}D_iW_0\overline{D_jW_0}=3|W_0|^{2}\,,
\eeq
and thus a no-scale scalar potential for the complex structure moduli and the axio-dilaton
\beq
V_{flux}=e^{K}\left( K^{S\bar S}\left|D_SW_0\right|^{2} + K^{a\bar{b}} D_a W_0 \overline{D_b W_0} \right)\,.
\eeq

This potential is positive semi-definite. Hence, every flux-induced isolated supersymmetric extremum for the axio-dilaton and the complex structure moduli has a positive-definite mass matrix, and is a true local minimum. For $W=W_0$ the K\"ahler moduli are flat directions. Volume moduli stabilization may now proceed by several methods.  KKLT generates SUSY AdS minima for the volumes from non-perturbative effects balancing off a small $W_0$. On the other hand, in LVS  an interplay between non-perturbative and perturbative effects generates non-SUSY AdS vacua at exponentially large volumes, or AdS/dS vacua at largish volumes in the K\"ahler uplifting scenario studied here. All these stabilization methods necessarily  proceed via breaking the no-scale structure. The question of full stability of a given scenario or concrete model rests then on the effect of the no-scale breaking contributions from volume stabilization to the mass matrix of the complex structure moduli and the axio-dilaton. One can show for both 
the LVS and the K\"ahler uplifting scenario, that all no-scale breaking terms are suppressed by an extra inverse power $1/{\cal V}$ of the volume compared to the flux-induced piece above~\cite{Balasubramanian:2005zx,Rummel:2011cd}. As the flux-induced piece is positive semi-definite and ${\cal O}(1/{\cal V}^{2})$, any negative term must come from no-scale breaking contributions which are ${\cal O}(1/{\cal V}^{3})$. Hence, any small shift of $S$ or one of the $U_a$ will see a positive ${\cal O}(1/{\cal V}^{2})$ increase in the scalar potential overwhelming any possible decreasing ${\cal O}(1/{\cal V}^{3})$ contribution from K\"ahler moduli stabilization. The only condition for this automatic separation of scales to work is stabilization of the over-all volume at largish values in the first place.

Thus, \emph{any} choice of flux producing an isolated SUSY extremum $D_SW=D_{U_a}W=0$ will be a minimum of the full potential once the LVS or K\"ahler uplifting generate a local minimum for the K\"ahler moduli at large volumes. Thus there is no need for a detailed model-by-model calculation of flux-induced mass matrix for the complex structure moduli and the axio-dilaton (a task practically unfeasible for typical values $h^{2,1}={\cal O}(100)$).

\section{A fully stabilized de Sitter vacuum of $\Pex$} \label{dS_sec}

In this section, we show that all geometric moduli of $\Pex$ can be stabilized in a metastable de Sitter vacuum. The stabilization of the two K\"ahler moduli via the interplay of non-perturbative effects in the superpotential and $\alpha'$ corrections in the K\"ahler potential will be discussed in section~\ref{Kahler_sec}. The analysis will single out values for the flux superpotential $W_0$, the string coupling $g_s$ and the one-loop determinant of gaugino-condensation $A$ which have to be realized to construct a de Sitter vacuum. We will demonstrate in section~\ref{CSex_sec} that the values of these parameters can be provided by explicitly solving a two dimensional subspace of the complex structure moduli space of $X_3$, hence providing an explicit construction of a de Sitter vacuum.

\subsection{K\"ahler uplifted de Sitter vacua} \label{Kahler_sec}

Realizing the brane and gauge flux setup discussed in sections~\ref{IIBperspectiveEx} and~\ref{FthEx_sec} there are two K\"ahler moduli $T_1 = \tau_1 + i \zeta_1$ and $T_2 = \tau_2 + i \zeta_2$ whose real components correspond to the volumes of the divisors $D_1$ and $D_5$, i.e.\ $\tau_1 \equiv \Vol_1$ and $\tau_2 \equiv \Vol_5$. The volume form of the three-fold eq.~\eqref{Volex} leaves us with the following K\"ahler potential for the K\"ahler moduli:
\begin{equation}
 K = -2 \log \left[ \frac{1}{\sqrt{12}} \left((T_1 +  \bar T_1) + \frac{1}{3} (T_2 + \bar T_2) \right)^{3/2} - \frac{1}{18} (T_2 + \bar T_2)^{3/2} + \frac{1}{2}\hat\xi(S,\bar S) \right]\,,
\end{equation}
where $\hat\xi(S,\bar S)$ is given in eq.~\eqref{xihatgen} with $\chi = 2(2-272) = -540$.

The superpotential arising from gaugino condensation of a $SU(24)$ and $SO(24)$ pure Super Yang Mills (SYM) on the D7-brane stacks wrapping $D_1$ and $D_5$ is given as
\begin{equation}
 W = W_0 + A_1\,e^{- \frac{2\pi}{24} T_1} + A_2\,e^{- \frac{2\pi}{22} T_2}\,,
\end{equation}
i.e. $a_1 = 2\pi/24$ and $a_2 = 2\pi /22$. We know that the one-loop determinants of the two gaugino condensates
are non-zero since $D_5$ is rigid and $D_1$ has been `rigidified' by flux.\footnote{The `rigidifying' flux on $D_1$ induces a correction to the gauge kinetic function of the form $T_1 \rightarrow T_1 - \frac{1}{2 g_s} \int_{D_1} \mathcal{F}_{D_1} \wedge \mathcal{F}_{D_1}$ which is non-zero since $\int_{D_1}  \mathcal{F}_{D_1} \wedge \mathcal{F}_{D_1} = -6$, see eq.~\eqref{intF1F1D1}. This can be absorbed in the one-loop determinant $A(S,U_a)$ since the correction does not depend on the $T_i$. We thank Gary Shiu for bringing this to our attention.} However, it is not know how to calculate the dependence of $A_1$ and $A_2$ on the complex structure moduli and the axio-dilaton. After stabilizing these moduli, $A_1$ and $A_2$ (as well as $W_0$) are flux-dependent parameters and can be treated as constant in the K\"ahler moduli stabilization process. In general, they will be of the same order as the VEVs of the complex structure moduli. Since there is a large number of flux parameters due to $h^{2,1} = 272$, it 
seems a reasonable assumption that one should be able to use the freedom in this sector to mildly tune $A_1$ and $A_2$ to a value desired for the stabilization of the K\"ahler moduli. Nevertheless, the unknown dependence of the one-loop determinants on the complex structure moduli remains the only point that cannot be addressed explicitly in our construction of a de Sitter vacuum.

The scalar potential of the moduli $T_1$, $T_2$ is
\begin{equation}
 V=e^{K} \left( K^{T_i\bar{T}_j} \left[ W_{T_i} \overline{W_{T_j}} + (W_{T_i} \cdot \overline{W K_{T_j}} + c.c) \right] + 3 \hat\xi \frac{\hat{\xi}^2+7\hat{\xi}\hat{\mathcal{V}}+\hat{\mathcal{V}}^2}{(\hat{\mathcal{V}}-\hat{\xi})(\hat{\xi}+2\hat{\mathcal{V}})^2} |W|^2 \right) \,.\label{VT1T2}
\end{equation}
Considering the dilaton $S$ and the complex structure moduli $U_a$ to be fixed by fluxes, which will be done explicitly in section~\ref{CSex_sec}, eq.~\eqref{VT1T2} is a function $V=V(T_1,T_2)$ with parameters $W_0$, $S$, $A_1$ and $A_2$. A de Sitter vacuum will only be obtained for certain values of these four parameters. In the following we consider the axio-dilaton to have a real VEV $s=\text{Re}(S)$, i.e.\ the axionic component is stabilized at zero VEV.

In~\cite{Rummel:2011cd} a sufficient condition on the compactification parameters was found which ensures that a de Sitter vacuum can be constructed via the method of K\"ahler uplifting. The condition applies when the manifold is of the swiss-cheese type. For a one parameter model this condition reads
\begin{equation}
 3.65 \lesssim \frac{-27 W_0 \hat\xi a^{3/2}}{64 \sqrt{2} \gamma A} \lesssim 3.89\,,
\label{suffcond}
\end{equation}
where $\gamma$ is related to the self-intersection number of the wrapped four cycle. For our brane setup on $\Pex$ the volume form is only approximately of the swiss-cheese type so that the analytic results of~\cite{Rummel:2011cd} cannot be applied directly. Nevertheless, we can study the scalar potential given in eq.~\eqref{VT1T2} numerically, searching for stable de Sitter minima for fixed numerical values of $W_0$, $S$, $A_1$ and $A_2$.

\begin{figure}[t!]
\centering
\includegraphics[width= 0.6\linewidth]{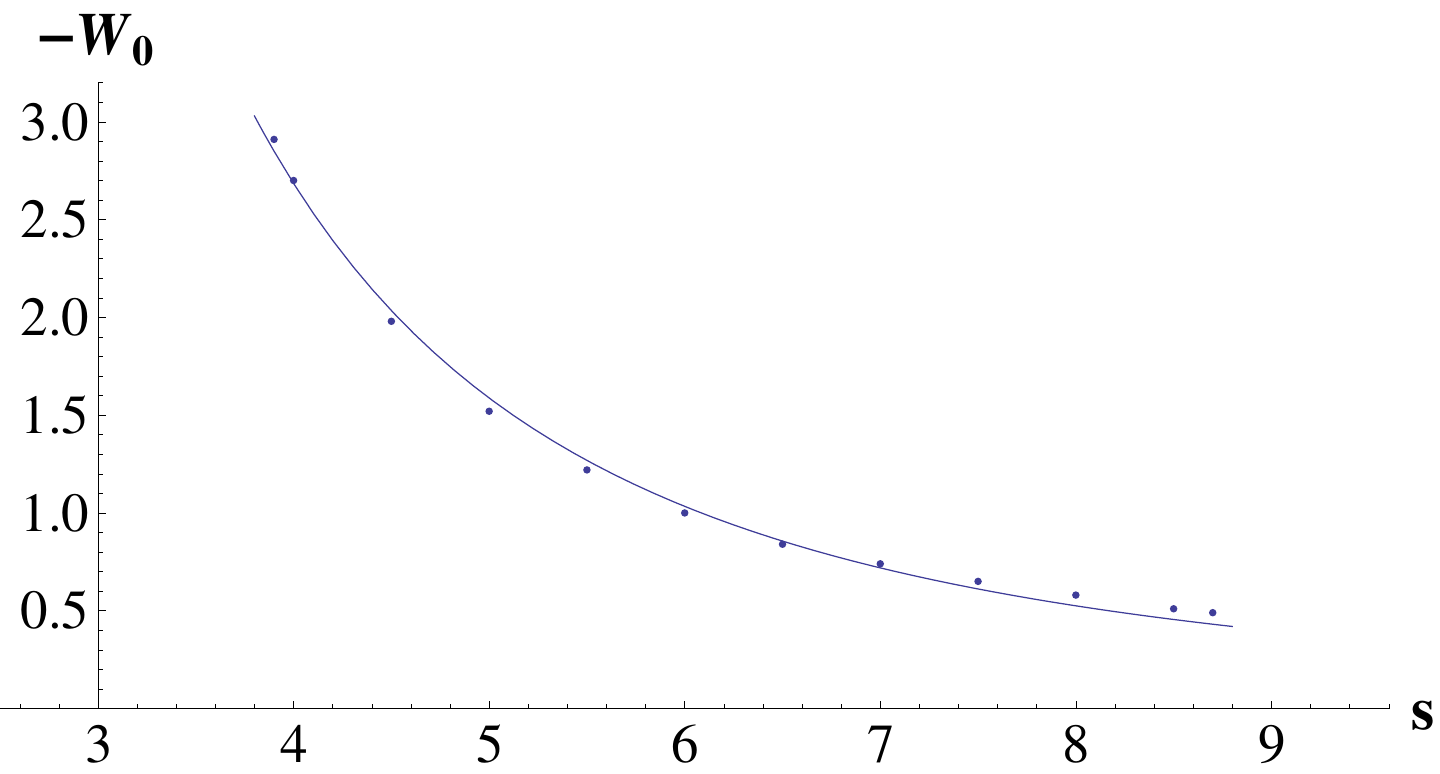}
\caption{The points represent $W_0, \,S$ pairs that allow a stable de Sitter solution of the potential given in eq.~\eqref{VT1T2} for $A_1=A_2=1$. The curve represents a fit $W_0 = C_1 s^{-C_2}$, with $C_1 = 70.2$ and $C_2 = 2.35$. Note that there is a small deviation from the one modulus case where $C_2=1.5$.}
\label{W0Splot}
\end{figure}

We find that there is indeed a subspace in this parameter space where the two K\"ahler moduli are stabilized in a stable de Sitter vacuum. For instance, keeping $A_1$ and $A_2$ constant we find that there is a curve in $W_0-S$ space that allows stable de Sitter vacua, see Figure~\ref{W0Splot}. Actually, it is a band rather than a curve, the lower bound of the band corresponding to Minkowski vacua and the upper bound corresponding to the minimum becoming an inflection point. For a single K\"ahler modulus this corresponds to the upper and lower bound in eq.~\eqref{suffcond}, respectively. Since the width of the band is rather small and we are interested in vacua with a small cosmological constant, we choose to display the lower bound in Figure~\ref{W0Splot}.

In the remainder of this section, let us study explicitly the following point in parameter space that realizes a de Sitter vacuum:
\begin{equation}
 W_0 = 0.812\,,\qquad s=6.99\,,\qquad A_1= 1.11\,,\qquad A_2=1.00\,.\label{parex}
\end{equation}
The choice of numerical values in eq.~\eqref{parex} is due to the solutions we find in the complex structure sector, see Section~\ref{CSex_sec}. $W_0$ and $s$ originate from this sector and $A_1$ and $A_2$ have to be chosen appropriately, invoking the constraint that generically they are of the same order as the VEVs of the complex structure moduli, in this case $\order(1)$ as we will see in section~\ref{CSex_sec}. 

We can justify both treating the instanton prefactors $A_i$ as effective constants, and mildly tuning them. Recent work has shown~\cite{Rummel:2011cd} that for large volume the mass scale of the K\"ahler moduli separates from the scale of the axio-dilaton and the complex structure moduli by one inverse power of the volume. This justifies replacing the complex structure moduli by their VEVs inside the 1-loop determinants, and allows us to parametrize these prefactors as effective constants. Moreover, we can clearly dial the VEVs of the complex structure moduli by availing ourselves of the exponentially large flux discretuum, which easily accounts for a potential mild tuning of the value of the 1-loop determinants.

\begin{figure}[t!]
\centering
\includegraphics[width= 0.49\linewidth]{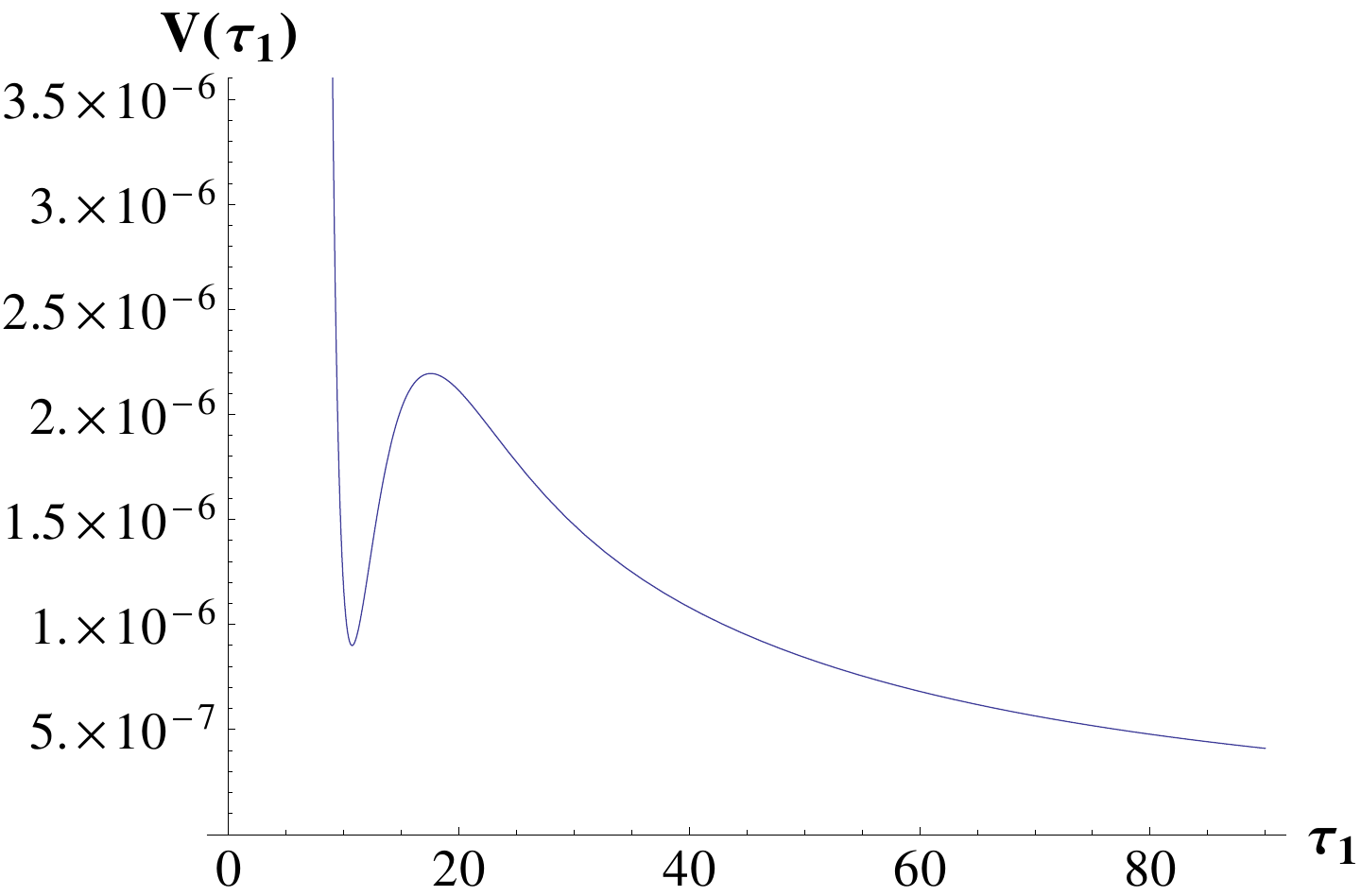}
\includegraphics[width= 0.49\linewidth]{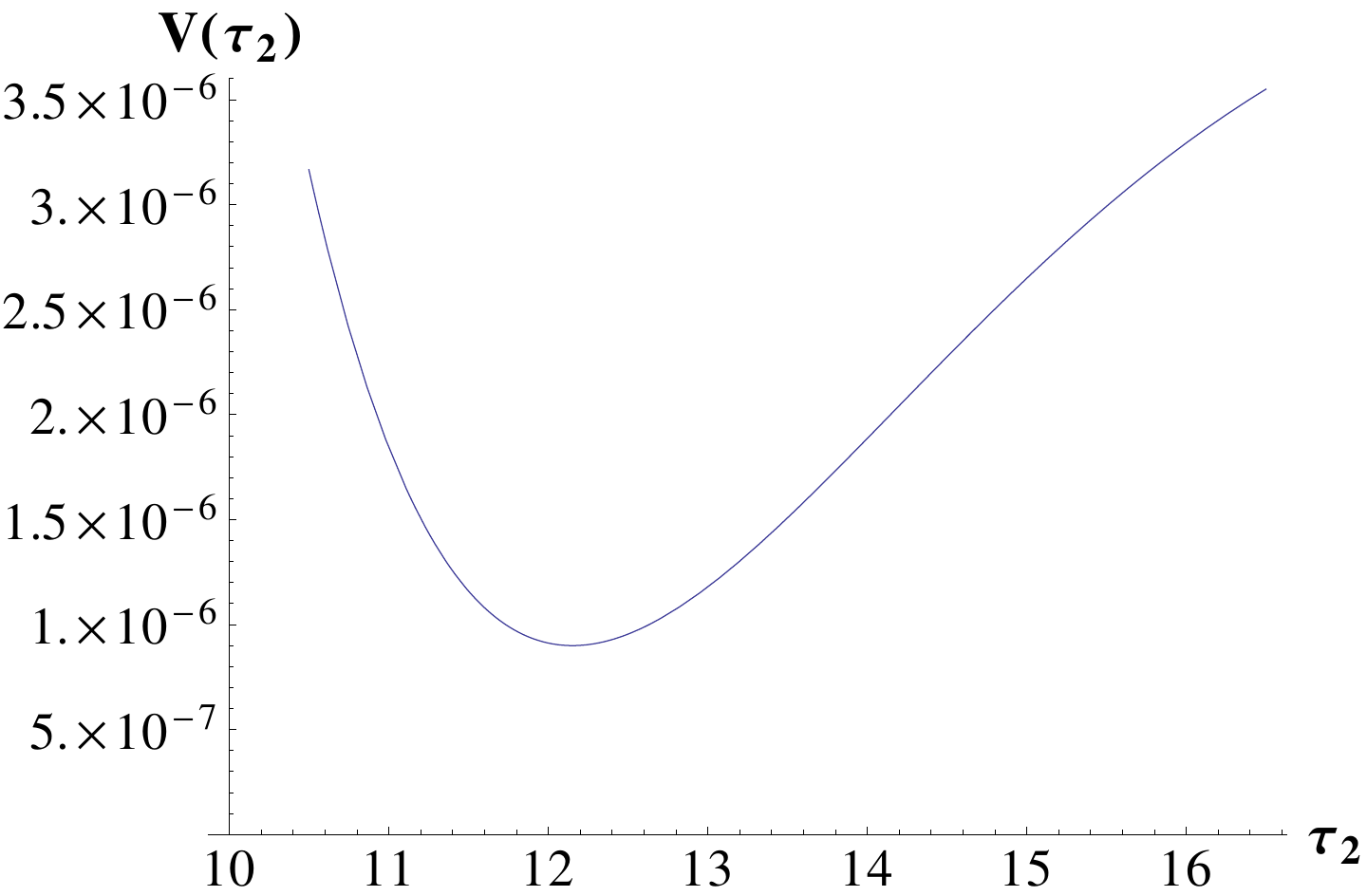}
\vskip 1mm
\includegraphics[width= 0.49\linewidth]{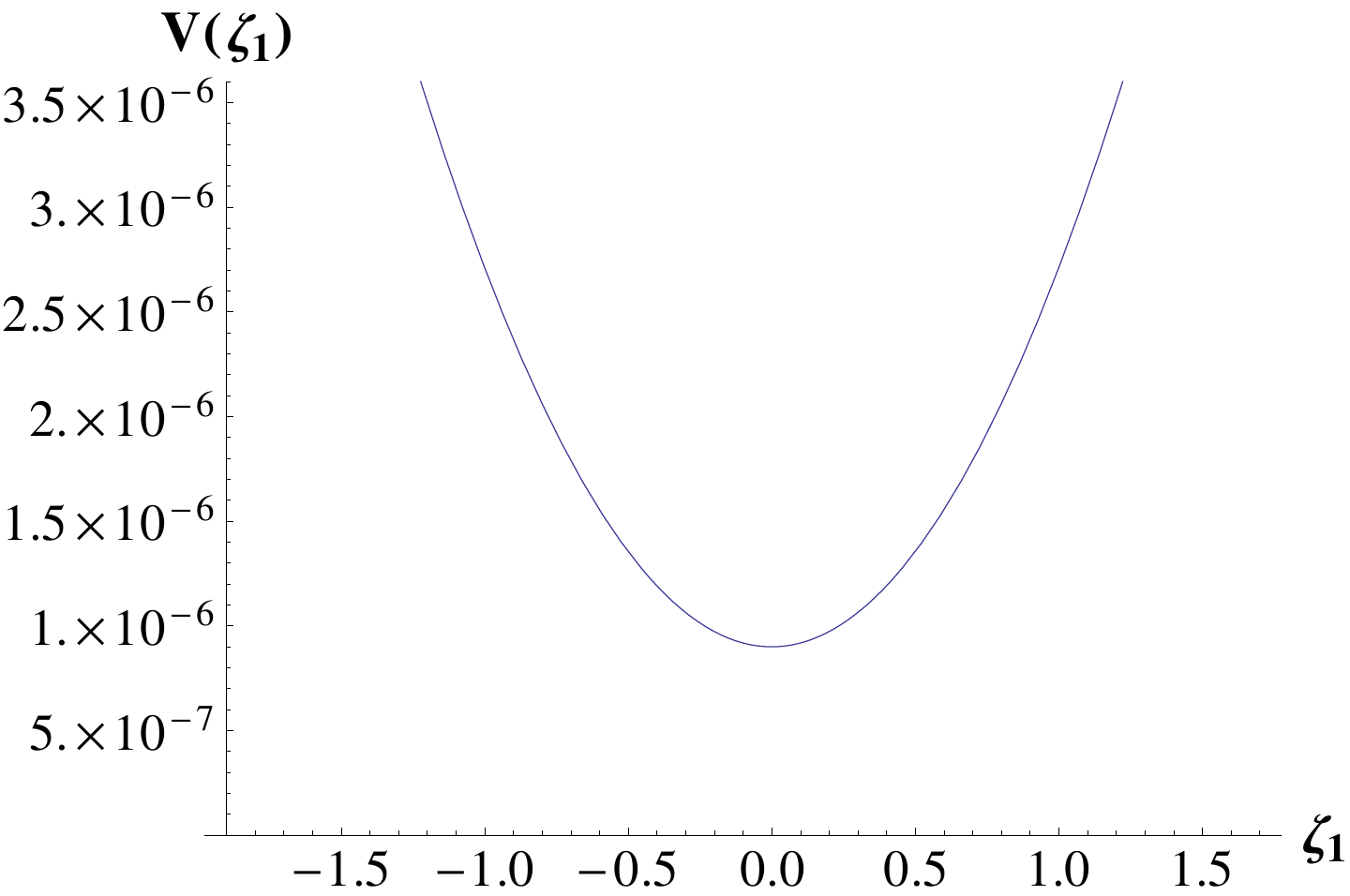}
\includegraphics[width= 0.49\linewidth]{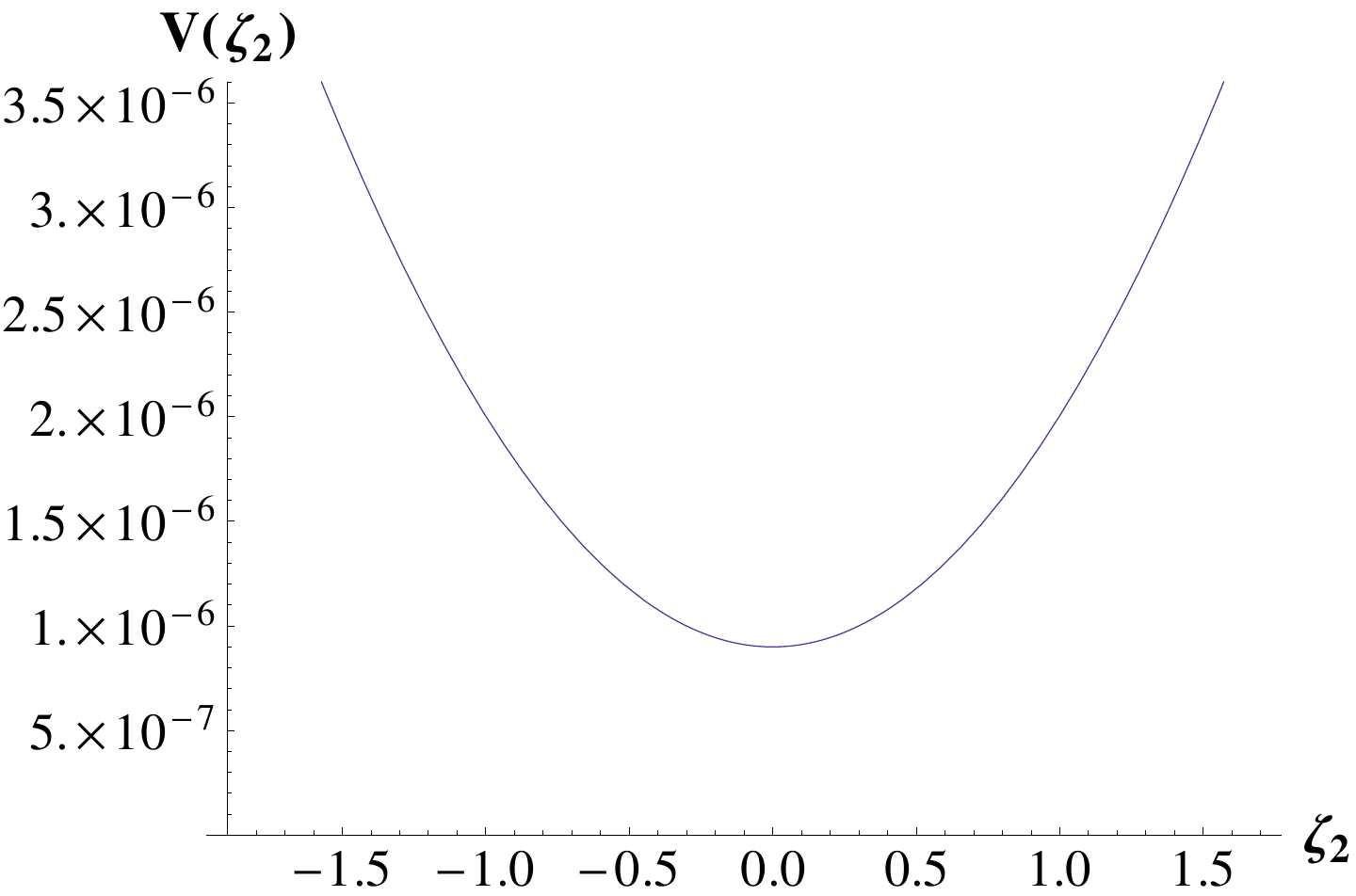}
\caption{The scalar potential $V(T_1,T_2)$ is a function of four real scalar fields. We show $V(T_1,T_2)$ as a single valued function with the other three fields evaluated at the minimum.}
\label{Vt1t2plot}
\end{figure}

The phenomenology of the model~\eqref{parex} is summarized in Table~\ref{modulimasses} and Figure~\ref{Vt1t2plot}. In particular, we note that the overall volume and the volume of the divisors $D_1$ and $D_5$ are stabilized at $\order(10-100)$. The not too large overall volume emerges from the fact that we have only realized an $N_1=24$ gauge group on $D_1$ which is actually lower than the critical gauge group rank $\sim 30$, as was discussed in section~\ref{lggr_sec}. Note that we were forced to choose the rank smaller than the maximal rank $N_{lg}=27$ in order to consistently incorporate the subtleties in the D7-brane configuration and construct a fully consistent model, see Section~\ref{IIBperspectiveEx}. Since models with a larger number of K\"ahler moduli allow in principle larger maximal gauge group rank, one may also realize larger overall volumes in these more complicated cases.

The K\"ahler moduli are stabilized inside the K\"ahler cone which corresponds to $t_1 > 3\,t_5$. Inverting the relation between the $t_i$ and the $\Vol_i$, eq.~\eqref{Volex}, we find $t_1 = 9.43$ and $t_5 = 1.50$ such that the K\"ahler cone condition is fulfilled.
\vskip 4mm
\begin{table}[t!]
\centering
  \begin{tabular}{c|c|c|c|c|c|c|c}
  $\langle T_1 \rangle$ & $\langle T_2 \rangle$ & $\langle \Vol \rangle$ & $m_{\tau_1}^2$ & $m_{\tau_2}^2$ & $m_{\zeta_1}^2$ & $m_{\zeta_2}^2$ & $m_{3/2}^2$\\
  \hline
  $10.76$ & $12.15$ & $51.97$ & $5.24 \cdot 10^{-9}$ & $4.55 \cdot 10^{-8}$ & $1.13 \cdot 10^{-7}$ & $6.40 \cdot 10^{-8}$ & $4.08 \cdot 10^{-7}$
  \end{tabular}
  \caption{VEVs and masses of the K\"ahler moduli. $m_{3/2}^2 = e^K |W|^2$ denotes the gravitino mass. For the calculation of the masses the prefactor $\langle e^{K_{\text{c.s.}}} \rangle \simeq 0.03$ from the complex structure moduli stabilization, see following section~\ref{CSex_sec}, has been taken into account. All masses are given in units of the Planck mass while volumes are given in units of $\alpha'$.}
  \label{modulimasses}
\end{table}

\subsection{Complex structure moduli} \label{CSex_sec}

In this section, we study the complex structure moduli space of $X_3$. In the end, we will present explicit RR and NS flux choices that stabilize the dilaton and complex structure moduli supersymmetrically such that $W_0$ and $s$ take the values in eq.~\eqref{parex}. Note that switching on RR and NS flux will not introduce Freed-Witten anomalies~\cite{Freed:1999vc} since the four-cycles $D_1$ and $D_5^{\rm fix}$ wrapped by D7-branes have no three-cycles (as $h^{1,0}(D_1) = h^{1,0}(D_5^{\rm fix}) = 0$, see table~\ref{chi0D1D5IIB}). Hence, any three-form of the threefold $X_3$ is pulled back to zero on the D7-branes, making the Freed-Witten constraint $H_3|_{D7}=0$ automatically satisfied.%
\footnote{Unfortunately we are not able to check the Freed-Witten anomaly cancellation due to $H_3$ for the Whitney brane. In fact, due to its singular world volume, the solution of this problem for a generic Whitney brane is so far unknown.}
Since it was shown in the previous section that the parameters~\eqref{parex} lead to a stable de Sitter vacuum in the K\"ahler moduli sector, $\Pex$ will provide us with an explicit example of such a vacuum with all geometric moduli stabilized. 

As was discussed in~\cite{Greene:1990ud,Candelas:1994hw,Denef:2004dm}, the 272 dimensional complex structure moduli space of $\Pex$ is symmetric under a $\Gamma = \mathbb{Z}_6 \times\mathbb{Z}_{18}$ action. 
Under $\Gamma$, only two complex structure moduli are left invariant. As was noted in~\cite{Denef:2004dm} it is sufficient to turn on fluxes only along the six $\Gamma$-invariant three-cycles to achieve $D_i W = 0$ for \textit{all} 272 complex structure moduli, and then to find a minimum of the positive definite tree-level no-scale scalar potential $V=\sum_i |D_i W|^2$. 
This is due to the fact that, for this invariant flux, the symmetry $\Gamma$ is realized at the level of the four-dimensional effective action.
Note that the restriction to flux on the $\Gamma$ invariant cycles is purely for simplicity, as the analysis of the complete 272 dimensional complex structure moduli space is practically not doable.

Once the symmetric fluxes are switched on, the K\"ahler potential and superpotential are
\begin{align}
 &K = - \log \left(-i\,\Pi^\dagger \cdot \Sigma \cdot \Pi \right) - \log \left(S+\bar S \right) \,,\label{Kcs}\\
 &W_0 = 2\pi(f-S\,h )\cdot \Pi\,,\label{W0cs}
\end{align}
where $\Pi^I$ is the vector of periods of the Calabi-Yau threefold holomorphic three-form $\Omega_3$ on a symplectic basis of three-cycles,
$\Sigma$ is the canonical symplectic matrix and $f=(f_1,\dots,f_6,0,...,0)$ and $h=(h_1,\dots,h_6,0,...,0)$ are the integer valued RR and NS flux quantum numbers (we have set to zero all the components along the $b_3-6$ non-invariant three-cycles. 

Let us explain why this flux vector generically provides a stable minimum of \textit{all} 272 complex structure moduli~\cite{Giryavets:2003vd,Denef:2004dm}. We first consider $D_{\tilde U_i} W_0 = 0$, where $\tilde U_i$ for $i=3,\dots,272$ denote the non-trivially transforming moduli under $\Gamma = \mathbb{Z}_{6}\times \mathbb{Z}_{18}$. The period vector is, at leading order, a polynomial function of the $\tilde U_i$. Furthermore, $\Pi$ cannot contain linear powers of the $\tilde U_i$, since these would not be invariant under $\Gamma$ and the period vector has to respect the symmetry of the complex structure moduli space. This information is sufficient to show
\begin{equation}
 W_{0,\tilde U_i} = K_{\tilde U_i} = 0 \quad \text{at} \quad \tilde U_i = 0 \quad \text{for} \quad i=3,\dots,272\,,
\end{equation}
since $W_{0,\tilde U_i}$ is a polynomial function which is at least linear in the $\tilde U_i$, see eq.~\eqref{W0cs} and $K_{\tilde U_i}$ is a rational function which is at least linear in the numerator in the $\tilde U_i$, see eq.~\eqref{Kcs}. Hence, $D_{\tilde U_i} W_0 = W_{0,\tilde U_i} + K_{\tilde U_i} W_0 =0$ at $\tilde U_i = 0$ for $i=3,\dots,272$.
This reduces the full set of conditions $D_iW=0$ $\forall i$ to the three equations
\begin{equation}
 D_{\phi}W |_{\tilde U_i = 0}=0 \qquad \mbox{for } \,\, \phi=S,U_1,U_2\,,
\end{equation}
where $U_1$ and $U_2$ are the two invariant complex structure moduli. 
This is equivalent to set $\tilde U_i = 0$ from the beginning and study the stabilization problem for the reduced case with two complex structure moduli, as we do in the following.

In~\cite{Candelas:1994hw}, the prepotential $\mathcal{G}$ for the two complex structure moduli $U_1 \equiv \nu_1 + i\, u_1 = \omega_1/\omega_0$ and $U_2 \equiv \nu_2 + i\, u_2 = \omega_2/\omega_0$ was derived via mirror symmetry in the large complex structure limit to be
\begin{equation}
 \mathcal{G}(\omega_0,\omega_1,\omega_2) = \xi  \omega _0^2+\frac{17 \omega _0 \omega _1}{4}+\frac{3 \omega _0 \omega _2}{2}+ \frac{9 \omega _1^2}{4}+ \frac{3 \omega _1 \omega _2}{2}-\frac{9 \omega _1^3+9 \omega _1^2 \omega _2+3 \omega _1 \omega _2^2}{6 \omega _0}\,,\label{prepotex}
\end{equation}
with $\xi=\frac{\zeta(3)\chi}{2(2\pi\,i)^3}\simeq -1.30843\,i$ determined by the Euler number $\chi$ of the Calabi-Yau. 

Eq.~\eqref{prepotex} receives instanton corrections which are given as
\begin{align}
 \mathcal{G}_{\text{inst.}}(q_1,q_2)=  \frac{1}{(2\pi\,i)^3}&\left( 540 q_1+\frac{1215 q_1^2}{2}+560 q_1^3+3 q_2-1080 q_1 q_2+143370 q_1^2 q_2 \right.\notag\\ & \left.-\frac{45 q_2^2}{2}+2700 q_1 q_2^2+\frac{244 q_2^3}{9} + \dots \right)\,,
\label{Ginst}
\end{align}
with $q_i = \exp{(2\pi\,i\,\omega_i)}$ and we have set $\omega_0=1$. The dots in eq.~\eqref{Ginst} denote higher powers in the $q_i$ which are suppressed in the large complex structure limit $\omega_i \gg 1$.

For a symplectic basis of three-cycles, the period vector is given by
\begin{align}\label{periodvec}
 \Pi=&(\partial_{\omega_0} \mathcal{G},\partial_{\omega_1} \mathcal{G},\partial_{\omega_2} \mathcal{G},\omega_0,\omega_1,\omega_2)\nonumber\\
 =& (2 \xi +\frac{17 U_1}{4}+\frac{3 U_1^3}{2}+\frac{3 U_2}{2}+\frac{3}{2} U_1^2 U_2+\frac{1}{2} U_1 U_2^2,\\ \notag&\frac{17}{4}+\frac{9 U_1}{2}-\frac{9 U_1^2}{2}+\frac{3 U_2}{2}-3 U_1 U_2-\frac{U_2^2}{2},\frac{3}{2}+\frac{3 U_1}{2}-\frac{3 U_1^2}{2}-U_1 U_2,1,U_1,U_2)\,.
\end{align}

Finally, the D3-charge induced by the RR and NS fluxes is given by
\begin{equation}
 Q_{D3}^{f,h} = \frac{1}{2}\, h\cdot \Sigma \cdot f = 66 \label{tadF3H3}\:.
\end{equation}

Our goal is to find flux quanta $f$ and $h$ that stabilize $S$, $U_1$ and $U_2$ supersymmetrically, i.e. $D_\phi W_0 = 0$ for $\phi = S,\,U_1,\,U_2$, with $\langle W_0\rangle$ and $\langle S\rangle$ suitable to perform the K\"ahler moduli stabilization in a de~Sitter vacuum.

To find stationary points we apply the strategy suggested in~\cite{Denef:2004dm}:\footnote{A detailed analysis of the flux solution space will be presented in~\cite{MMRW:2012toappear}. Among others, one can use a recently developed method, called numerical polynomial homotopy continuation method, which can find all the stationary points of a given potential having polynomial-like nonlinearity~\cite{Mehta:2012wk}.}
\begin{itemize}
 \item Neglecting instanton corrections to the prepotential and approximating $|\xi| \simeq 1.30843\dots$ by an approximate rational value, for instance $|\xi| = 13/10$ we solve the system of equations
\begin{equation}
 0 =(W_0,D_{S}W_0,D_{U_1}W_0,D_{U_2}W_0)_{(\text{no inst.})}\,,
\end{equation}
 for the flux quanta $f$ and $h$ setting the VEVs $S$, $U_1$ and $U_2$ to fixed rational values. Furthermore, we require $f$ and $h$ to fulfill the tadpole constraint found in eq.~\eqref{totalD3charge}, $Q_{D3}^{f,h} \leq 104$ or $Q_{D3}^{f,h} \leq 96$. This amounts to solving a linear equation of the form $A\cdot (f,h) = 0$ with $A\in \mathbb{Q}^{8\times12}$ for $f$ and $h$, respecting the tadpole constraint.
 \item The flux solution for $f$ and $h$ is inserted into
\begin{equation}
 0 = (D_{S}W_0,D_{U_1}W_0,D_{U_2}W_0)_{(\text{inst.})}\,,\label{CSsolvenoinst}
\end{equation}
 where this time the instanton corrections to the prepotential eq.~\eqref{Ginst} are taken into account in calculating $D_i W_0$ and the constant $\xi$ is set to its exact value. This generates shifts in the VEVs of $S$, $U_1$ and $U_2$ from their original rational values. Also the superpotential $W_0$ may be shifted from its zero value, eq.~\eqref{CSsolvenoinst} to a non-vanishing value. If the resulting values for $S$ and $W_0$ are suitable for the K\"ahler moduli stabilization, we have constructed a de Sitter vacuum.
\end{itemize}

Let us present the solution that provides the parameters of our example given in eq.~\eqref{parex}. The flux vector
\begin{equation}
 (f;h)=(-16, 0, 0, 0, -4, -2; 0, 0, 2, -8, -3, 0)\,,
\end{equation}
induces a D3 charge $Q_{D3}^{f,h} = 66$. 
Since this does not saturate the negative contribution to the total D3 tadpole, one has to switch on additional trivial gauge flux on the brane stacks or introduce a number of D3-branes to obtain an overall vanishing D3 charge.

The VEVs of the moduli and superpotential are
\begin{equation}
 \langle S\rangle=6.99 \,,\quad \langle U_1\rangle=1.01 \,,\quad \langle U_2\rangle=0.967 \,,\quad \langle |W_0|\rangle=0.812 \,.
\end{equation}
A posteriori, we see that the chosen values for $A_1$ and $A_2$ in eq.~\eqref{parex} are indeed of the same order as the VEVs of the complex structure moduli. Furthermore, the assumption of working in the large complex structure limit is valid since the VEVs of $U_1$ and $U_2$ fulfill the condition that the instanton corrections are small, $U_2 \gg 1/6$ and $U_1 > 1$~\cite{Denef:2004dm}. Finally, we calculate the masses of the complex structure and dilaton moduli to 0th order from the tree level potential $V=e^K K^{a\bar{b}} D_a W_0 \overline{D_{b} W_0}$ for $a,b=U_1,U_2,S$ in table~\ref{modulimassesCS}.
\begin{table}[ht!]
\centering
  \begin{tabular}{c|c|c|c|c|c}
  $m_{u_1}^2$ & $m_{u_2}^2$ & $m_{s}^2$ & $m_{\nu_1}^2$ & $m_{\nu_2}^2$ & $m_{\sigma}^2$\\
  \hline
  $0.24$ & $1.8 \cdot 10^{-4}$ & $5.6 \cdot 10^{-6}$ & $0.24$ & $1.8 \cdot 10^{-4}$ & $5.7 \cdot 10^{-6}$
  \end{tabular}
  \caption{Masses of the real complex structure moduli $u_1$, $u_2$ the dilaton $s$ and their corresponding axion fields $\nu_1$, $\nu_2$ and $\sigma$. For the calculation of the masses the prefactor $\langle e^{K_{\text{K\"ahler}}} \rangle \simeq 3.7 \cdot 10^{-4}$ from the K\"ahler moduli stabilization, see table~\ref{modulimasses}, has been taken into account.}
  \label{modulimassesCS}
\end{table}
A posteriori, we verify that there is indeed a separation of scales, i.e.\ the complex structure moduli and the dilaton are stabilized at a mass scale roughly two orders of magnitude higher than the K\"ahler moduli.

To conclude this section, we have explicitly constructed a de Sitter vacuum with all geometric moduli stabilized on $X_3$. The stabilization of the two K\"ahler moduli, the dilaton and two complex structure moduli has been carried out explicitly while the remaining 270 complex structure moduli are stabilized according to general arguments.

\section{Conclusions}\label{conclusions_sec}

We discussed the construction of explicit global models in a type IIB context, which exhibit the dynamics of K\"ahler uplifting. In this mechanism the interplay of gaugino condensation on 7-branes and the leading ${\cal O}(\alpha'^{3})$-correction fixes the K\"ahler moduli in a SUSY breaking minimum, after three-form flux has supersymmetrically stabilized the complex structure moduli and the axio-dilaton. The vacuum energy of this minimum can be dialed from AdS to dS by adjusting the flux-induced Gukov-Vafa-Witten superpotential using the flux discretuum. Both SUSY breaking and lifting to dS are driven by an F-term of the K\"ahler moduli sector arising from the presence of the $\alpha'$-correction of the K\"ahler potential. Thus the dS uplift is realized entirely by
the geometric closed string moduli. This was the motivation for trying to construct a fully explicit consistent global model including explicit flux choice and complex structure moduli stabilization.

In K\"ahler uplifted dS vacua the CY volume scales as $\Vol \propto N^{3/2}$ with $N$ the rank
of the condensing gauge group living on the 7-brane stack wrapping the
large four-cycle. Moreover, the scale
of K\"ahler moduli stabilization and the resulting K\"ahler moduli
masses 
are suppressed  by an additional ${\cal O}(1/\Vol)$ compared to the scale of flux-induced complex structure moduli stabilization. Hence, we searched for a large gauge group rank to obtain a large volume.

We considered models which can be easily uplifted to F-theory compactifications on
elliptically fibered CY fourfolds (embedded in toric spaces). We focused on 
models that possess Sen's weak coupling limit since we use the leading $\alpha'$ correction which is not understood for generic points in the F-theory moduli space. Such models have the characteristic feature that the orientifold plane has to be in a homology class of high degree in order to obtain a singularity of large rank $N$.

We checked the consistency conditions to have a globally defined construction, e.g. that the toric base of the elliptically fibered fourfold should be free of singularities of any kind  
as well as the Calabi-Yau threefold hypersurface in the weak coupling limit. 
We found that in general this turns out to be a severe constraint when one tries to increase the class of the orientifold by choosing appropriately the weights defining the toric variety.

We then made sure that the volume was of swiss cheese type $\Vol \sim \Vol_1^{3/2} - \Vol_i^{3/2}$, or at least approximately swiss cheese, e.g $\Vol \sim (\Vol_1 + \Vol_i)^{3/2} - \Vol_i^{3/2}$. Then one can manufacture a large overall volume by making $\Vol_1$ large by enforcing a large gauge group rank on the corresponding divisor $D_1$. We ensure that other large rank stacks wrapping some divisors $D_{i\not=1}$ (possibly enforced by imposing large rank on $D_1$)
did not destroy the large volume approximation.

Finally, we checked that the number of the neutral and charged zero modes could be put to zero, such that the gaugino condensation contribution to the superpotential is non-zero. To do this, the gauge flux on the brane stack has to be chosen appropriately: On the one hand it must be non-zero to `rigidify' the wrapped divisor (if this is not rigid). On the other hand, it should be possible to tune the flux such that it does not generate additional zero modes in the form of chiral matter, charged under the condensing gauge group.

We studied constraints on large gauge group rank by discussing Kreuzer-Skarke models and hypersurfaces in toric varieties. For the subclass of threefolds with an elliptic F-theory lift ($\sim 10^{5}$ models) we extracted the distribution of the largest-rank gauge group as a function of the number of K\"ahler moduli $h^{1,1}$.

Choosing $\Pex$, which has $h^{1,1}=2$ and $h^{2,1}=272$, as our explicit example we constructed large-rank singularities on a choice of two divisors, and analyzed the consistency constraints both in the type IIB Sen limit, and from the F-theory perspective. The emerging situation for $\Pex$ looks summarily as follows: We construct an $Sp(24)$ singularity on the `large' divisor $D_1$, which is rigidified by gauge flux, breaking $Sp(24)$ to $SU(24)$. The presence of the $Sp(24)$ stack forces an $SO(24)$ singularity on the second divisor $D_5$, which already is rigid. The gauge flux can be tuned such that no further zero-modes are generated. In F-theory, the gaugino condensation superpotential is related to the superpotential generated by the M5-instantons wrapping the 
exceptional divisors $E_i$ in the resolved fourfold. In the considered case, we found that the exceptional divisors resolving the $Sp(24)$ singularity satisfy $\chi_0(E_i)\geq 1$, that is the necessary condition in the presence of fluxes such that the wrapped M5-instantons contribute to the superpotential.

From the general results for supersymmetric flux stabilization we know, that at the no-scale level the resulting scalar potential is positive semi-definite, which yields full stability of the complex structure sector and the dilaton once fluxes fix them at an isolated supersymmetric point. As K\"ahler moduli stabilization via K\"ahler uplifting proceeds by breaking the no-scale structure at sub-leading order in the volume (like LVS), the stability of the flux-stabilized complex structure sector extends to the full model.
We then analyzed the scalar potential that stabilizes the K\"ahler moduli. This singled out a band in $g_s$ - $W_0$ plane where one finds de Sitter vacua. Here $W_0$ denotes the VEV of the Gukov-Vafa-Witten superpotential arising from supersymmetric flux stabilization of the complex structure moduli. The overall volume of the Calabi-Yau threefold was  determined by the data of the construction to be $\Vol \sim 52$.

The complex structure moduli space of $\Pex$ possesses a high-order discrete symmetry $\Gamma$. We only considered three-form fluxes that respect this symmetry. As a consequence, all the $h^{2,1}-2$ non-invariant complex structure moduli are stabilized. The prepotential of the remaining two complex structure moduli is known via mirror symmetry, and this enables us to stabilize all $h^{2,1}$ moduli explicitly. It is this fact that  in the end allowed us to construct a completely stabilized de Sitter vacuum. 

Finally, we gave an explicit flux choice which stabilizes the
axio-dilaton and the two $\Gamma$-invariant complex structure moduli
at the right VEVs and value for $W_0$ for the K\"ahler stabilization
of the explicit construction to proceed into a metastable dS
vacuum. In summary, we gave a construction of an example for dS space
in string theory which we believe to be explicit and complete within
the limits of existing knowledge. 

\acknowledgments We thank Andreas Braun, Volker Braun, Michele Cicoli, Andres Collinucci,
Frederik Denef, I\~naki Garc\'ia-Etxebarria, Thomas
Grimm, James Halverson, Arthur Hebecker, Shamit Kachru, Magdalena Larfors, Christoph Mayrhofer, Danny Mart\'inez Pedrera, Liam McAllister, Raffaele Savelli, Gary Shiu, Washington Taylor, Stefan Theisen, Timo Weigand, and Timm Wrase for valuable, and enlightening discussions, and useful comments. This work was supported by the Impuls und Vernetzungsfond of the Helmholtz Association of German Research Centers under grant HZ-NG-603, the German Science Foundation (DFG) within the Collaborative Research Center 676 "Particles, Strings and the Early Universe" and 
the Research Training Group 1670.

\appendix

\section{Models from two line weight systems} \label{twolineweight_sec}

In this appendix we want to discuss the constraints on building large gauge groups in Calabi-Yau threefolds that are complete intersections of hypersurfaces in a projective ambient space that is characterized by a weight system of two lines. This restriction is for simplicity. In all the approaches discussed below we find that going to arbitrarily high gauge groups would correspond to introducing singularities in the threefold of different kinds. In this sense, this appendix is meant as a summary of what can go wrong when one tries to build large gauge groups in a singularity free compact Calabi-Yau.

\subsection{One Hypersurface Calabi-Yaus}

In this section we will analyze the threefold that is the hypersurface
\begin{equation}
 \xi^2 - P_{2n+6,4} = 0\,,\label{xionehypapp}
\end{equation}
in the ambient space
\begin{equation}
X_4^{\text{amb}} :\quad
\begin{array}{cccccc}
u_1 & u_2 & u_3 & u_4 & u_5 & \xi\\ 
\hline 
1 & 1 & 1 & n & 0 & n+3\\
0 & 0 & 0 & 1 & 1 & 2\\
\end{array}\quad .
\label{WSPnApp}
\end{equation}
Notice that in the case $n>6$ a $u_5^2$ term factors out from the polynomial $P_{2n+6,4}$ due to the imposed scalings. The case $n=6$, that was studied in detail in section~\ref{IIBperspectiveEx}, has the largest $n$ where the enforced factorization of $P_{2n+6,4}$ is only linear in $u_5$. In the linear case, the O7 plane splits into two planes that for $n=6$ are not intersecting. When $n>6$ the picture is more complicated as there are intersecting $O7$ planes, which produce orbifold singularities on the Calabi-Yau threefold.

Notice that the constraint on $n$ is less restrictive if one gives up the weak coupling limit and goes to a strongly coupled F-theory compactification. For example, when $n=18$ the following factorization on $u_5$ in enforced on the Tate polynomials $a_i$:
\begin{align}
\begin{aligned}
 &a_1 = u_5^1 A_{21,1}\,,\\ 
 &a_2 = u_5^2 A_{42,2}\,,\\
 &a_3 = u_5^3 A_{63,3}\,,\\
 &a_4 = u_5^4 A_{84,4}\,,\\
 &a_6 = u_5^5 A_{126,7}\,,
\end{aligned}
\end{align}
which corresponds to an $E_8$ singularity as one can look up in~\cite{Bershadsky:1996nh}. For $n>18$ there is a factorization $a_6 = u_5^m A_{18+6n,12-m}$ with $m\geq 6$. This singularity cannot be resolved according to the Tate procedure and hence $n=18$ is the maximum value we can obtain in a strongly coupled F-theory setting. See also~\cite{Denef:2008wq} for a discussion of this limit on $n$ derived in the Weierstrass parametrization of the fourfold.

Finally, note that the class of the orientifold is $[O7]=21[D_1]+2[D_5]$ in the $n=18$ case while it is $[O7]=9[D_1]+2[D_5]$ in the weakly coupled $n=6$ case. Hence, in principle one could construct a much larger gauge group on $D_1$ if one gives up the weak coupling limit. If the leading $\alpha'$-correction of the K\"ahler potential were known in the strong coupling regime one could use this to construct K\"ahler uplifted de Sitter vacua where the large volume regime is more easily achieved.

\subsection{Complete Intersections}

One can also study complete intersections in higher dimensional toric varieties. For instance, take a threefold that is defined by two equations in a five dimensional ambient toric variety. This corresponds to a base which is a hypersurface in a four-dimensional ambient toric variety \cite{Collinucci:2009uh}. In the remainder of this section, we demonstrate in an example that if one wants to have a non-singular base of the corresponding F-theory uplift and an approximately swiss-cheese intersection form, this constraints how large the coefficients in $\Kbar$ and hence in $[O7]$ can become. Also in other examples that we considered, trying to enlarge $\Kbar$ at some point introduces singularities in the base or the double cover Calabi-Yau threefold.

As a working example, take the base manifold $B_{nl}$ that can be described as a hypersurface $P_{m,2}(u_i)=0$ with positive integer $m$ in the projective space
\begin{equation}
B_{nl}^{\text{amb}}\,:\,\,\,
\begin{array}{cccccc}
u_1 & u_2 & u_3 & u_4 & u_5 & u_6\\ 
\hline 
1 & 1 & 1 & n & l & 0\\
0 & 0 & 0 & 1 & 1 & 1\\
\end{array}\,\,.
\label{basenapp}
\end{equation}
The ambient four complex dimensional toric variety eq.~\eqref{basenapp} can be interpreted as a $\mathbb{CP}^2$ fibration over a $\mathbb{CP}^2$ with integer twists $n$ and $l$ which we choose to be positive. Note that the degrees of $P$ have been chosen such that the anti-canonical bundle of $B_{nl}$ is $\Kbar = (3+n+l-m)[D_1] + [D_6]$. In the previous section, the factorization of $u_5^2$ for $n>6$ in the $\xi^2-P_{2n+6,4}=0$ equation in eq.~\eqref{xionehypapp} was due to the relation $\Kbar \sim 2[D_5]$. We want to avoid this bound on $n$ here by making $\Kbar \sim [D_6]$.

Factoring out those coordinates that are sections of the bundle $\mathcal{L}_2$ defined by the second line in eq.~\eqref{basenapp}, the hypersurface equation is given as
\begin{align}
 P_{m,2}(u_i) &= u_6^2 P_m + u_6 u_5 P_{m-l} + u_5^2 P_{m-2l} + u_6 u_4 P_{m-n} + u_5 u_4 P_{m-l-n} + u_4^2 P_{m-2n}\,,\notag\\
\label{Pm2fac}
\end{align}
where the $P_i$ are sections only in the bundle $\mathcal{L}_1$ corresponding to the first line of eq.~\eqref{basenapp}. To avoid factorization of eq.~\eqref{Pm2fac} in $u_6$ and hence a singularity in the base, we have to impose 
\begin{equation}
2l \leq m\,. 
\label{2lleqm}
\end{equation}
Note that the roles of $n$ and $l$ can be exchanged since they just correspond to a redefinition of the coordinates $u_4$ and $u_5$. If we wanted $l$ to obtain large values compared to $m$ and $n$ we should impose $2n \leq m$. $\Kbar$ obtains a large scaling in $\mathcal{L}_1$ as long as $n$ or $l$ are allowed to be large. In the following, we denote divisors in the ambient space corresponding to $u_i=0$ by $\tilde D_i$ while we symbolize the pullback on the base manifold by $D_i$.

\subsubsection*{Singularities of the ambient space}

The fan of the toric variety $B_{nl}^{\text{amb}}$ can be generated by the lattice vectors
\begin{equation}
v_1= \left(\begin{array}{c}-n\\-l\\-1\\-1 \end{array}\right),\,
v_2= \left(\begin{array}{c} 0\\0\\1\\0 \end{array}\right),\,
v_3= \left(\begin{array}{c} 0\\0\\0\\1 \end{array}\right),\,
v_4= \left(\begin{array}{c} 1\\0\\0\\0 \end{array}\right),\,
v_5= \left(\begin{array}{c} 0\\1\\0\\0 \end{array}\right),\,
v_6= \left(\begin{array}{c} -1\\-1\\0\\0 \end{array}\right)\,.
\label{v1v6}
\end{equation}
$B_{nl}^{\text{amb}}$ is non-singular if all cones of the fan that generates the toric variety are generated by a subset of a lattice basis of the (in this case) four dimensional lattice. This is the case if all combinations $\text{span}_\mathbb{Z}\{v_i, v_j, v_k, v_l\}$ for $i,j,k,l \in \{1,..,6\}$ that are part of the toric variety form a basis of the lattice space. For the fan spanned by the $v_i$'s of eq.~\eqref{v1v6} the combinations that do not span the lattice are
\begin{align}
\begin{aligned}
 &\{v_1,v_2,v_3,v_4\}\,,\{v_1,v_2,v_3,v_4\}\,,\{v_1,v_2,v_3,v_4\}\,,\\
 &\{v_1,v_4,v_5,v_6\}\,,\{v_2,v_4,v_5,v_6\}\,, \{v_3,v_4,v_5,v_6\}\,.\label{dangerouscombs}
\end{aligned}
\end{align}
We will show in the next paragraph that $\tilde D_1 \tilde D_2 \tilde D_3$ and $\tilde D_4 \tilde D_5 \tilde D_6$ are part of the Stanley-Reisner ideal and hence the cones corresponding to eq.~\eqref{dangerouscombs} do not belong to the toric variety which is thus free of singularities.

\subsubsection*{Intersections in the ambient space}

We will now derive the quadruple intersections in the ambient space by analyzing the system of equations defined by the GLSM~\cite{Witten:1993yc} of eq.~\eqref{basenapp}
\begin{align}
\begin{aligned}
 |x_1|^2+|x_2|^2+|x_3|^2+n|x_4|^2+l|x_5|^2+|x_6|^2 &= \xi_1 > 0\,,\\
 |x_4|^2+|x_5|^2+|x_6|^2 &= \xi_2 > 0\,.
\label{GLSMamb}
\end{aligned}
\end{align}
A GLSM is a description of a toric variety where the D-flatness conditions of eq.~\eqref{GLSMamb} can be thought of as gauge fixing the absolute values of the rescalings in a complex projective space, while the $U(1)$ symmetry associated to the phase of the rescalings remains to be divided out. We immediately see from eq.~\eqref{GLSMamb} that $\tilde D_4 \tilde D_5 \tilde D_6 = 0$. Consider now the intersections
\begin{align}
\begin{aligned}
&\tilde D_1 \tilde D_2 \tilde D_3 \tilde D_4\,:\,\, |x_5|^2 = \frac{\xi_1}{l}\,,\,\, |x_6|^2 = \frac{1}{l}(l \xi_2-\xi_1)\,,\\
&\tilde D_1 \tilde D_2 \tilde D_3 \tilde D_5\,:\,\, |x_4|^2 = \frac{\xi_1}{n}\,,\,\, |x_6|^2 = \frac{1}{n}(n \xi_2-\xi_1)\,,\\
&\tilde D_1 \tilde D_2 \tilde D_3 \tilde D_6\,:\,\, |x_4|^2 = \frac{1}{l-n}(l \xi_2 - \xi_1)\,,\,\, |x_6|^2 = \frac{1}{l-n}(\xi_1 - n \xi_2)\,.
\end{aligned}
\end{align}
Independently of $n$ and $l$ these intersections vanish if
\begin{equation}
 \xi_1 > n \xi_2 \qquad \text{and} \qquad \xi_1 > l \xi_2\,.
\label{ineqssuff}
\end{equation}
In this case, $\tilde D_1 \tilde D_2 \tilde D_3$ and $\tilde D_4 \tilde D_5 \tilde D_6$ are elements of the Stanley Reisner ideal and hence the ambient space is singularity free. That the inequalities eq.~\eqref{ineqssuff} are indeed fullfilled will be shown in the next section.

To calculate all quadruple intersections it is sufficient to calculate the intersections for a basis of divisors, we pick $\tilde D_1$ and $\tilde D_6$. All other intersections can be calculated using the equivalences
\begin{equation}
\tilde D_1 = \tilde D_2 = \tilde D_3\,, \,\, \tilde D_4 = n \tilde D_1 + \tilde D_6\,, \,\, \tilde D_5 = l \tilde D_1 + \tilde D_6\,.
\label{divequivamb}
\end{equation}
Under the assumption eq.~\eqref{ineqssuff} we can show $\tilde D_1^2 \tilde D_5 \tilde D_6 = 1$. Successive use of the relations eq.~\eqref{divequivamb} gives us the quadruple intersections
\begin{align}
\begin{aligned}
&\tilde D_1^2 \tilde D_6^2 = 1\,,\qquad
\tilde D_1 \tilde D_6^3 = -n-l\,,\qquad
\tilde D_6^3 = n^2 + l(n+l)\,.\label{intsamb}
\end{aligned}
\end{align}

\subsubsection*{Mori- and K\"ahler cone of the ambient space}

The Mori cone, i.e.\ the set of two-cycles that generate the full set of all two-cycle classes with holomorphic representatives, is spanned by $C_1 = \tilde D_1 \tilde D_6^2$ and $C_2 = \tilde D_1^2 \tilde D_6$. The weight matrix
\begin{equation}
 \tilde D_i \cdot C_a =
\left( \begin{array}{cccccc}
1 & 1 & 1 & -l & -n & -n-l\\
0 & 0 & 0 & 1 & 1 & 1\\
\end{array}\right) \,\,,
\end{equation}
can be obtained from the original GLSM description eq.~\eqref{basenapp} by adding the second line of eq.~\eqref{basenapp} multiplied by $-n-l$ to the first line. This corresponds to the transformation of the Fayet-Iliopoulos terms of the GLSM according to
\begin{equation}
 (\xi_1,\xi_2) \rightarrow ( \xi_1', \xi_2') = (\xi_1 - (n+l) \xi_2,\xi_2)\,.
\end{equation}
For a basis of divisors $K_i$ that is dual to the $C_a$, i.e.\ $K_i C_a = \delta_{ia}$ the K\"ahler form can be parametrized as $J= \xi_i' K_i$. In this case, $K_1 = \tilde D_1$ and $K_2 = (n+l)\tilde D_1 + \tilde D_6$. The K\"ahler cone, i.e.\ the space of all two forms with
\begin{equation}
 \int_{C_i} J > 0\,,
\end{equation}
for all $C_i$ in the Mori cone, is then simply given by
\begin{equation}
 \xi_1' > 0, \,\,\, \xi_2' > 0 \qquad \Leftrightarrow \qquad \xi_1 - (n+l) \xi_2 > 0,\,\,\,\xi_2 > 0\,.
\label{xiprimeineqs}
\end{equation}
The inequalities in eq.~\eqref{xiprimeineqs} a posteriori justify the quadruple intersections calculated in eq.~\eqref{ineqssuff}.

\subsubsection*{Intersections in the base and volume form}

The intersections of the base divisors $D_i$ are calculated from those in the ambient space via
\begin{equation}
\kappa_{ijk}=\int_{B_{nl}} D_i D_j D_k = \int_{B_{nl}^{\text{amb}}} (m \tilde D_1 + 2 \tilde D_6) \tilde D_i \tilde D_j \tilde D_k\,.
\end{equation}
Using eq.~\eqref{intsamb} we thus find
\begin{align}
\begin{aligned}
&D_1^3=0\,,\\
&D_1^2 D_6=2\,,\\
&D_1 D_6^2= m - 2(l+n)\,,\\
&D_6^3= - m(l+n)+2(n^2+l(l+n))\,.\label{intsbasetemp}
\end{aligned}
\end{align}
For the purpose of moduli stabilization the volume of the Calabi-Yau\footnote{Note that the relation between the volume of the base and the volume of the Calabi-Yau could be more complicated as it was for instance the case for the threefold $X_3$ that we considered in section~\ref{IIBperspectiveEx}.}
\begin{equation}
 \Vol = 2 \Vol_{B_{nl}} = \frac{1}{3} \int_{B_{nl}} J^3 = \frac{1}{3} \int_{B_{nl}} \left(\xi_1 D_1 + \xi_2 D_6 \right)^3\,,
\end{equation}
should be 'approximately swiss cheese', i.e.\ of the form
\begin{equation}
 \Vol \sim (a V_{D_1} + b V_{D_6})^{3/2} - c V_{D_6}^{3/2} \qquad \text{with} \,\, a,c > 0\,.
\end{equation}
As was discussed in~\cite{MayrhoferThesis}, this can only be arranged if the two dimensional matrix
\begin{align}
 &A_{ij} = \kappa_{ijk} a^k\, \qquad \text{for} \qquad i,j,k=1,2\,,
\end{align}
with triple intersections $\kappa_{ijk}$ and $a^k \in \mathbb{Z}$ defining the divisor $a^1 D_1 + a^2 D_6$ can be brought into a special form: There is exactly one non-vanishing eigenvalue while the eigenvector corresponding to the vanishing eigenvalue is $(0,1)$. From the matrix entries
\begin{align}\begin{aligned}
A_{11}&=2 a_2\,,\\
A_{12}&=A_{21} = 2 a_1+a_2 (m+2 (-l-n))\,,\\
A_{22}&=a_1 (m+2 (-l-n))+a_2 \left(m (-l-n)+2 \left(n^2+l (l+n)\right)\right)\,,
\end{aligned}\end{align}
we see that these conditions can only be met if
\begin{equation}
m = 2l,\,\, a^1 = n\, a^2\, \qquad \text{or} \qquad m = 2n,\,\, a^1 = l\, a^2\,.
\label{m2la1na2}
\end{equation}
Since we are interested in large $n$ having already imposed $2l\leq m$ in eq.~\eqref{2lleqm} we stick to the first option in eq.~\eqref{m2la1na2} and furthermore choose $a^2 = 1$. Now the volume can be shown to be given by
\begin{equation}
 \Vol =\frac{2}{3n}\left[ (n V_{D_1} + V_{D_6})^{3/2} - V_{D_6}^{3/2} \right]\,.
\end{equation}
The anti-canonical bundle of $B_{nl}$ becomes
\begin{equation}
 \Kbar = (3+n-l)[D_1]+[D_6]\,.\label{KbarappA}
\end{equation}
Now going back to eq.~\eqref{Pm2fac} we see that the only way we can get a larger class than $3[D_1]$ in $\Kbar$ is by demanding $u_4^2 P_{m-2n} = u_4^2 P_{2(l-n)}=0$. This would however leave us with a singular base since along the curve $\{u_5=u_6=0\}$, which is not in the SR-ideal, we would have $P_{m,2} = d P_{m,2} = 0$. Hence, we find the constraint $n-l\leq 0$ and we did not succeed in constructing a large class of $[D_1]$ in $\Kbar$, as one can see from eq.~\eqref{KbarappA}.

\section{The $Sp(N)$ resolved fourfold} 

\subsection{Geometry of the $Sp(N)$ resolved fourfold} \label{SpNgeom}

In this section, following the procedure used in~\cite{Collinucci:2010gz,Collinucci:2012as}, we resolve the $Sp(N)$ singularity on a divisor $\hat{D}$ for a general base. $\hat{D}$ is defined by an equation $p_{\hat{D}}=0$, where $p_{\hat{D}}$ is a polynomial of the base coordinates with proper degrees.\footnote{We will make no attempts to resolve the $SO(N)$ singularity (for a discussion of an $SO(10)$ resolution, see~\cite{Tatar:2012tm}.) From the point of view of gaugino condensation in our example the resolution is not necessary since we know that $D_5^{\rm fix}$ is rigid in the type IIB Calabi-Yau threefold, Table~\ref{chi0D1D5IIB}. Resolving the $SO(N)$ singularity would be necessary if we wanted to calculate the D3 tadpole from the F-theory perspective, as we would require the Euler number of the completely resolved fourfold.} 
                                                                                                                                                                                                                                                                                                                                                                                                                                                                                                                                                        
\begin{figure}[h!]
\centering
\includegraphics[width= 0.65\linewidth]{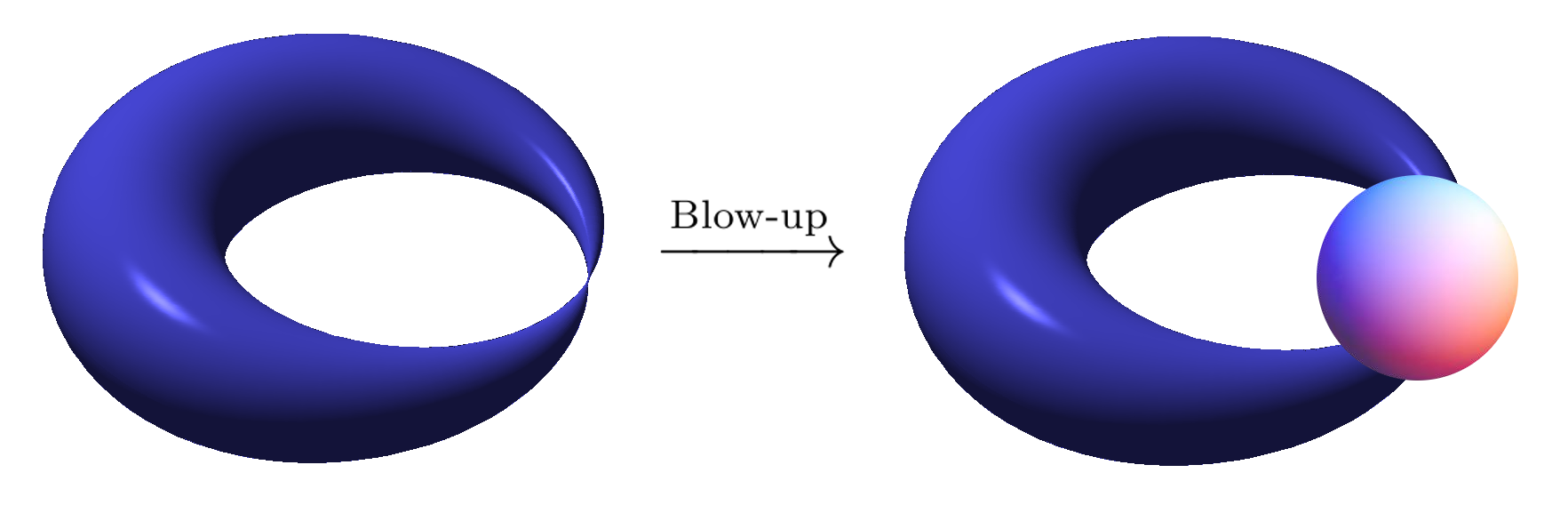}
\caption{Visualization of a blown up $Sp(1)$ singularity.}
\label{fig_Sp1blowup}
\end{figure}

The elliptic fibration becoming singular can be visualized by a pinching of the fibered torus, i.e.\ one of the torus cycles shrinks to zero size. The singularity is resolved via a series of blow-ups, where the blow-ups are such that they reproduce on the fiber the Dynkin diagram of the gauge group singularity that was originally imposed on the fiber. More precisely, the pullback to the base of the double intersections of the blown-up divisors build the Cartan matrix entries of the gauge group. For instance, the resolution of an $Sp(1)$ singularity is visualized in Figure~\ref{fig_Sp1blowup}. 

These blown-up divisors can also be formalized by means of toric geometry~\cite{Collinucci:2010gz,Collinucci:2012as}. To resolve an $Sp(N)$ singularity, $N$ new coordinates $v_{2i-1}$, $i=1,..,N$, are introduced with projective scaling relations 
\begin{equation}
X_6^{\text{amb}}\,:\,\,\,
\begin{array}{c|cccccccc}
\sigma & X & Y & Z & v_1 & v_3 & \cdots & v_{2N-3} & v_{2N-1}\\ 
\hline 
0 & 2 & 3 & 1 & 0 & 0 & \cdots & 0 & 0\\ 
1 & 1 & 1 & 0 & -1 & 0 & \cdots & 0 & 0\\ 
1 & 2 & 2 & 0 & 0 & -1 & \cdots & 0 & 0\\ 
\vdots & \vdots & \vdots & \vdots & \vdots & \vdots & \ddots & \vdots & \vdots\\ 
1 & N-1 & N-1 & 0 & 0 & 0 & \cdots & -1 & 0\\ 
1 & N & N & 0 & 0 & 0 & \cdots & 0 & -1
\end{array}\,\,.
\label{toric6fold}
\end{equation}
The resolved fourfold is embedded into the ambient sixfold $X_6^{\text{amb}}$~\eqref{toric6fold} by the two equations:
\begin{equation}
\begin{aligned}
Y^{\text{res}}\,:\,\,
\begin{cases}
 &Y\left(Y + a_1 X Z + a_{3,N} \Pi_{i=0}^N v_{2i-1}^{N-i} Z^3 \right)\\
 &= X^3 \Pi_{i=0}^N v_{2i-1}^{i} + a_2 X^2 Z^2 + a_{4,N} \Pi_{i=0}^N v_{2i-1}^{N-i} X Z^4 + a_{6,2N} \left(\Pi_{i=0}^N v_{2i-1}^{N-i} \right)^2 Z^6\,,\\
 &p_{\hat{D}}=\Pi_{i=0}^N v_{2i-1}\,.\\
\end{cases}
\end{aligned}
\end{equation}
where we have defined $v_{-1}\equiv\sigma$.

The vanishing of the new coordinates defines the exceptional divisors  $E_{2i-1}: \,\,\{v_{2i-1}=0\}$ and $E_{-1} \equiv [\sigma]$.
As far as divisor classes are concerned there are the following equivalences:
\begin{equation}
[\sigma]  = \hat{D} - \sum_{i=1}^N E_{2i-1}\,,\quad
[X] = 2([Z]+\Kbar) - \sum_{i=1}^N i\, E_{2i-1}\,,\quad
[Y] = 3([Z]+\Kbar) - \sum_{i=1}^N i\, E_{2i-1}\,.\label{YZK}
\end{equation}
where $[X],[Y],[Z]$ are the homology classes of the divisors $\{X=0\},\{Y=0\},\{Z=0\}$. With abuse of notation we call $\hat{D}$ both the four-cycle $\{p_{\hat{D}}=0\}$ on the base and its uplift to the Calabi-Yau fourfold, i.e.\ a six-cycle that is an elliptic fibration over the locus $\{p_{\hat{D}}=0\}$ on the base.

\subsection{Intersections of exceptional divisors from the Stanley-Reisner ideal}\label{sec_intsSpN}

The Stanley-Reisner (SR) ideal of the fourfold with a resolved $Sp(N)$ singularity is
\begin{equation}
 SR_{Sp(N)} = \{X Y Z, v_{2i-1} Z_{|i=1,..,N}, v_{2i-1} X_{|i=0,..,N-1}, v_{2i-1} v_{2j-1 \,| i,j=0,..,N;\, j-i>1}\}\,.
\label{SRSpN}
\end{equation}
This follows from the SR ideal of the ambient sixfold given in \cite{Collinucci:2012as}, restricted to the fourfold. One can directly deduce eq.~\eqref{SRSpN} from the fact that the SR ideal is the union of all sets
\begin{equation}
 Z_{I}=\{(z_1,..,z_n) | z_j=0 \,\forall j \in I\}
\end{equation}
for which there is no cone, such that all the one-cones $\rho_j$ associated to the homogeneous coordinate $z_j$ with $j\in I$ lie in one cone. Let us consider the case of $Sp(2)$ from where it is straightforward to deduce eq.~\eqref{SRSpN}. The SR ideal reads
\begin{equation}
 SR_{Sp(2)} = \{X Y Z, v_{1} Z, v_{3} Z, \sigma X, v_1 X, \sigma v_3\}\,.
\label{SRSp2}
\end{equation}
This can be obtained via the following polytope construction: From the scaling relations eq.~\eqref{toric6fold} one can make a canonical choice of one-cones $\rho_i$~\cite{Candelas:1997eh,Krause:2011xj} for the coordinate set $\{X,Y,Z,v_{2i-1\,|i=0,..,N}\}$:
\begin{equation}
x= \left(\begin{array}{c}-1\\0\\ \underline{0} \end{array}\right),\,
y= \left(\begin{array}{c}0\\-1\\ \underline{0} \end{array}\right),\,
z= \left(\begin{array}{c}2\\3\\ \underline{0} \end{array}\right),\,
v_{2i-1}= \left(\begin{array}{c}2-i\\3-i\\ \underline{v} \end{array}\right)\,,
\end{equation}
for $i=0,..,N$.

\begin{figure}[h!]
\centering
\includegraphics[width= 0.32\linewidth]{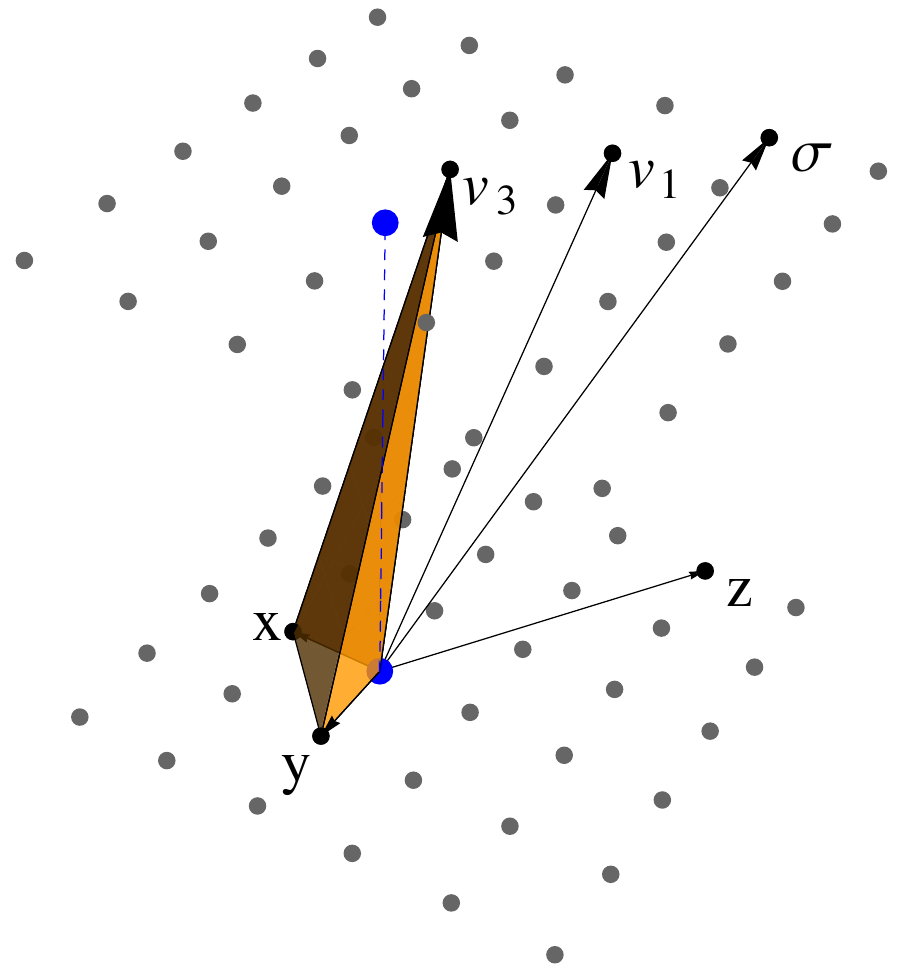}
\includegraphics[width= 0.32\linewidth]{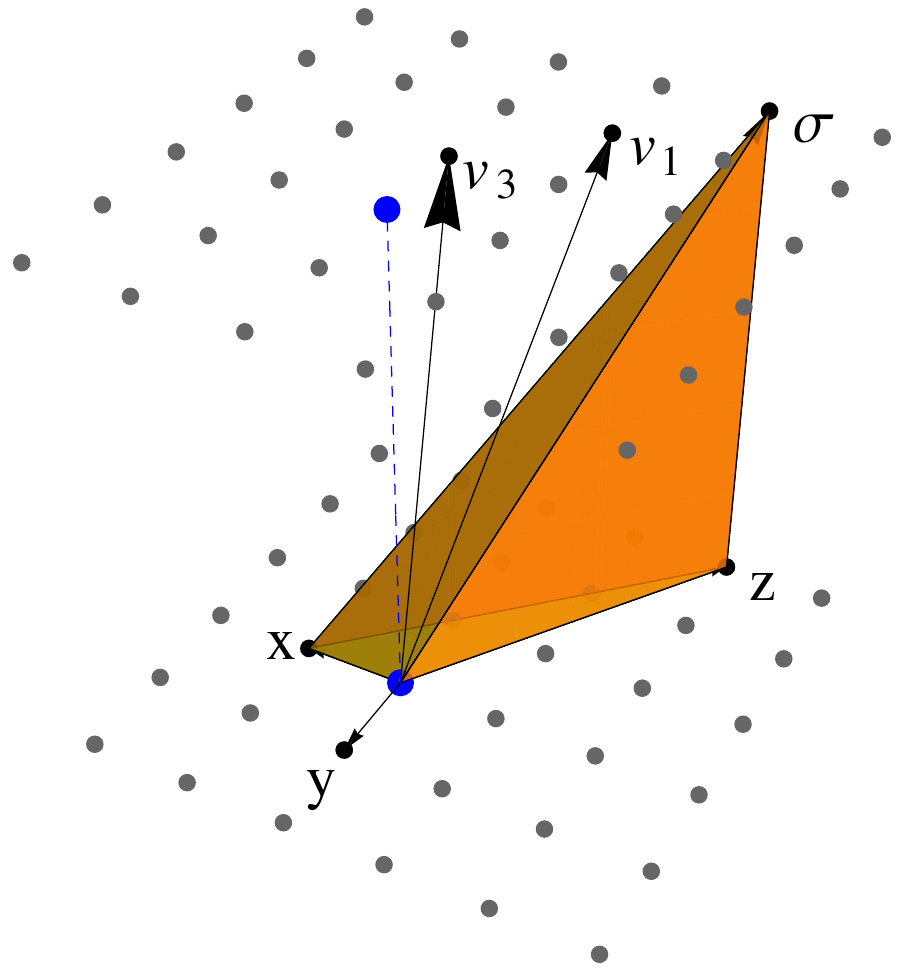}
\includegraphics[width= 0.32\linewidth]{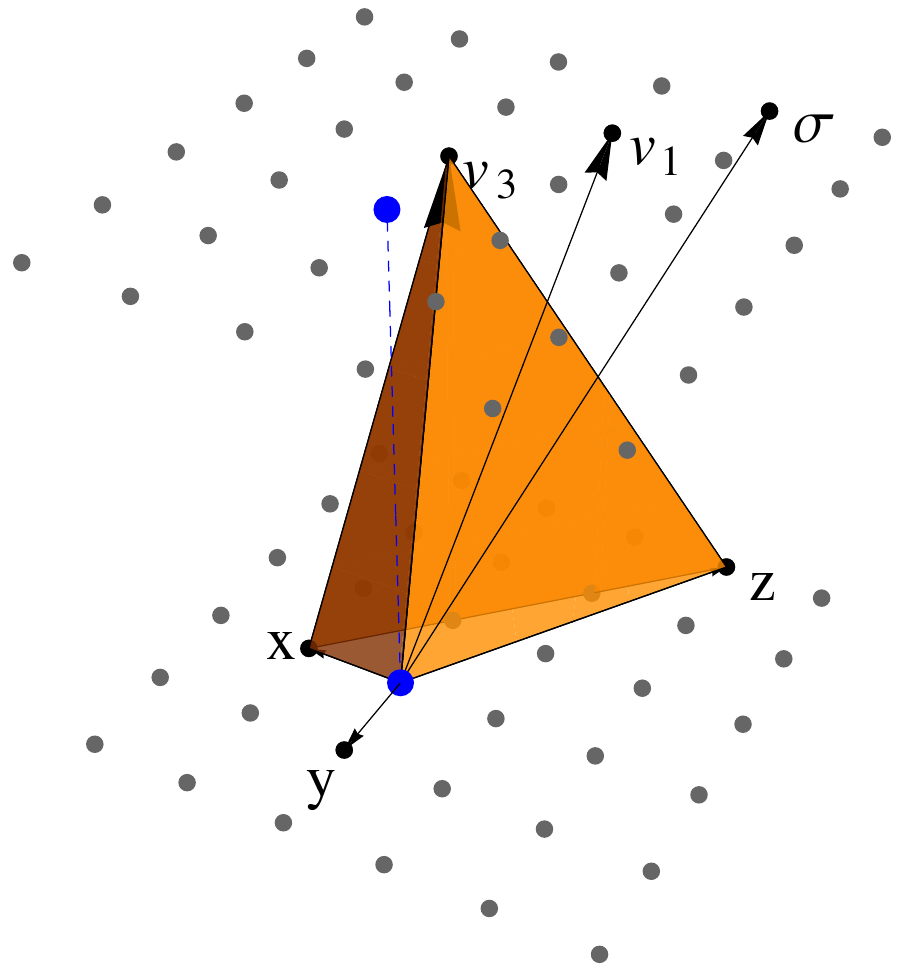}
\caption{3D projection of the fan of the $Sp(2)$ resolution manifold for the subset of coordinates $\{X,Y,Z,\sigma,v_1,v_3\}$. The top layer of grey lattice points corresponds to the projection $\underline{v}\rightarrow 1$ while the bottom layer to $\underline{0}\rightarrow 0$. The blue point indicates the origin. In the first plot from the LHS, we see the cone spanned by $\{x,y,v_3\}$ is such that the one-cones $x$, $y$ and $z$ can never lie in one cone and hence $XYZ$ is an element of the SR ideal. In the second plot, we see that the cone spanned by $\{x,z,\sigma\}$ is such that $z$, $v_1$ and $z$, $v_3$ respectively can never lie in one cone. In the third plot we see that the cone spanned by $\{x,z,v_3\}$ forces $X \sigma$ and $X v_1$ to lie in the SR ideal. $\sigma v_3$ is an element of the SR ideal because $v_1$ lies on a line that connects them and hence they can never lie in one cone. These are all possible elements of the SR ideal, notice that for example $y$ and $v_1$ lie in one cone: $\{y,v_1,v_3\}
$.}
\label{fig_polyhedraSpN}
\end{figure}

Mapping $\underline{0}\rightarrow 0$ and $\underline{v}\rightarrow 1$ we can draw a three dimensional projection of the fan which we show in Figure~\ref{fig_polyhedraSpN} for the case of $Sp(2)$. Using the definition of the SR ideal one can read of its elements. For a more detailed explanation see the caption of Figure~\ref{fig_polyhedraSpN}.

We now want to calculate the double intersections of the exceptional divisors $E_{2i-1} E_{2j-1}$ from the SR ideal~\eqref{SRSpN} using in principal the same strategy as was presented in~\cite{Krause:2012yh} to obtain the double intersections of the exceptional divisors for the $SU(N)$-resolution manifold for $N=2,..,5$. In the $Sp(N)$ case we consider here this analysis is more straightforward than in the $SU(N)$ case and can actually be used to obtain the intersections for an arbitrary $Sp(N)$ resolution manifold. This, in turn will be used to derive in the next section a formula for the arithmetic genus for the $Sp(N)$ resolution manifold with arbitrary $N$.

The most general form of the double intersections is
\begin{equation}
 E_{2i-1} E_{2j-1} = C_{ij}\, \hat{D} ([Z]+\Kbar) + d_m E_{2m-1} \hat{D} + k_m E_{2m-1} \Kbar\,,
\end{equation}
where $\hat{D}$ is the base divisor $\{p_{\hat{D}}=0\}$ where the $Sp(N)$ singularity is located. 
The $C_{ij}$ have to be the entries of the Cartan matrix of $Sp(N)$ times two, since $Sp(N)$ is not simply-laced. The coefficients $d_m$ and $k_m$ can be extracted from the SR ideal and the relations~\eqref{YZK}. For example $[\sigma][X] = 0$ implies:
\be
\left(\hat{D} - \sum_{k=1}^N E_{2k-1} \right)\left(2([Z]+\Kbar)-\sum_{i=1}^N i\, E_{2i-1} \right) = 0\,.
\ee

In general, we have to solve $2N-1$ linear equations
\begin{align}
\begin{aligned}
E_{2i-1} [X] = 0& \qquad \text{for} \,\, i=0,..,N-1\,,\\
[\sigma] E_{2i-1} = 0& \qquad \text{for} \,\, i=2,..,N\,.
\end{aligned}
\end{align}
for $2N-1$ non-zero intersections $E_{2i-1\,|i=1,..,N}^2$, $E_{2i-1}E_{2i+1\,|i=1,..,N-1}$. These equations read
\begin{align}
\begin{aligned}
 &\sum_{j=2}^{N-1} j\, E_{2j-1}(E_{2j-3}+E_{2j-1}+E_{2j+1})+ N E_{2N-1}(E_{2N-3}+E_{2N-1})\\ &+E_1(E_1+E_3)= -2 \hat{D} ([Z]+\Kbar) + \hat{D} \sum_{j=2}^{N-1} j\, E_{2j-1} + 2 \Kbar \sum_{j=2}^{N-1} E_{2j-1}\,,\\
 &E_1(E_1+2E_3)=2 E_1 \Kbar\,,\\
 &E_{2i-1} \left[ (i-1)E_{2i-3} + i E_{2i-1} + (i+1)E_{2i+1} \right] = 2 E_{2i-1} \Kbar;\,\, i=2,..,N-1\,,\\
 &E_{2i-1} \left[ \hat{D} - E_{2i-3} - E_{2i-1} - E_{2i+1} \right] = 0;\,\, i=2,..,N-1\,,\\
&E_{2N-1} \left[ \hat{D} - E_{2N-3} - E_{2N-1} \right] = 0\,.\label{}
\end{aligned}
\end{align}

It can be shown by mathematical induction $\forall N$ that these linear equations are solved by the following non-zero double intersections:
\begin{align}
\begin{aligned}\label{SpNint}
 E_{2i-1}E_{2i+1} = &2 \hat{D} ([Z]+\Kbar) - 2 \Kbar \sum_{k=i+1}^N E_{2k-1} - \hat{D} \sum_{k=1}^i k\, E_{2k-1};\,\, i=1,..,N-1\,,\\
 E_{2i-1}^2 = &-4 \hat{D}([Z]+\Kbar) +\Kbar \left(4 \sum_{k=i+1}^N E_{2k-1} +2 E_{2i-1} \right)\\ &+ \hat{D}\left( 2 \sum_{k=i+1}^{i-1} k\, E_{2k-1} + (i+1) E_{2i-1} \right);\,\, i=1,..,N-1\,,\\
 E_{2N-1}^2 = &-2 \hat{D} ([Z]+\Kbar) + 2 \Kbar E_{2N-1} + \hat{D} \left( \sum_{k=1}^{N-1} k\, E_{2k-1} + E_{2N-1} \right)\,.
\end{aligned}
\end{align}

\subsection{Arithmetic genus of divisors}\label{Sec:ArithmGen}

The arithmetic genus $\chi_0$ of a divisor $D$ in a fourfold $X_4^{\text{res}}$ is given by
\begin{align}
\begin{aligned}
 \chi_0(D) &= \frac{1}{24} \int_D c_1(D) c_2(D)\\
 &= \frac{1}{24} \int_{X_4^{\text{res}}} (-D^2)\left( c_2(X_4^{\text{res}}) + D^2 \right)\,.
\label{arithmeticgenusgen}
\end{aligned}
\end{align}
To calculate the second Chern class $c_2(X_4^{\text{res}})$ over a
base $B_3$ with toric divisors $\hat{D}_i$ note that the total Chern
class is given by
\begin{align}
\begin{aligned}
 c(X_4^{\text{res}}) &= c_{\text{fib.}}\,\prod_{\hat{D}_i\neq \hat{D}}(1+\hat{D}_i)(1+[\sigma])(1+E_1)..(1+E_{2N-1})\\
 &= c_{\text{fib.}}\,\prod_{\hat{D}_i\neq \hat{D}}(1+\hat{D}_i)(1+\sum_{j=0}^N E_{2i-1} + \sum_{0\leq k < l}^N E_{2k-1}E_{2l-1})\\
 &= c_{\text{fib.}}\,(1+\Kbar+c_2(B)+ \hat{D} \sum_{j=1}^N E_{2i-1} - \sum_{1\leq k < l}^N E_{2k-1}E_{2l-1}+..)\,,
\end{aligned}
\end{align}
where the dots denote terms that are at least triple intersections and 
\begin{equation}
 c_{\text{fib.}} = \frac{(1+[X])(1+[Y])(1+[Z])}{1 + 2 [Y]}\,.
\end{equation}
Using the relations~\eqref{YZK} and the intersection formula eq.~\eqref{SpNint} it is then straightforward to show
\begin{equation}
 c(X_4^{\text{res}}) = 1 + c_2(B) + 12 \Kbar [Z] + 11 \Kbar^2 - 7 \Kbar \sum_{i=1}^N i\, E_{2i-1} + \hat{D} \sum_{i=1}^N i^2 E_{2i-1} + ..\,,
\label{cchangeapp}
\end{equation}
where the dots denote the third and higher Chern classes. The arithmetic genus of various divisors can then be calculated from the definition eq.~\eqref{arithmeticgenusgen} using again the intersection formula eq.~\eqref{SpNint} and $E_{2i-1} [Z]=0_{|i=1,..,N}$. We list the arithmetic genus of the divisors that are of interest for us:
\begin{align}\begin{aligned}\label{chi0vNapp}
 \chi_0(E_{2i-1}) &= \frac{1}{6} \int_{B_3} \hat{D} \left[c_2(B)+\Kbar^2 + 2 \hat{D}^2\right],\,\, i=1,..,N-1\,,\\
 \chi_0(E_{2N-1}) &= \frac{1}{12} \int_{B_3} \hat{D} \left[c_2(B)+\Kbar^2 - 3 \Kbar \hat{D} + 2 \hat{D}^2\right]\,. 
\end{aligned}\end{align}
One can check that 
\begin{equation}\label{chi0rel}
 \chi_0(E_{2N-1})  = \chi_0(\hat{D}) \qquad \mbox{and} \qquad  \chi_0(E_{2i-1})  = \chi_0(D)\,, \,\,\,\, i=1,...,N-1\,,
\end{equation}
where $\hat{D}$ is the divisor $\{p_{\hat{D}}=0\}$ on the base manifold $B_3$ and $D$ is its double cover in $X_3$. The relations \eqref{chi0rel} can be expected, by considering that the exceptional divisors $E_{2i-1}$ are $\mathbb{P}^1$ fibrations over a divisor on the base manifold.\footnote{In particular, the elliptic fiber splits on top of $\{p_{\hat{D}}=0\}$ into a collection of $\mathbb{P}^1$s whose mutual intersection is encoded into the Dynkin diagram of the corresponding ADE-singularity. When the singularity is `non-split', i.e.\ some of the $\mathbb{P}^1$s are interchanged by a monodromy when going around the base divisor, then the gauge group is not of A-D-E type.
The monodromies are reflected by symmetries of the ADE Dynkin diagram, leading to the diagram of the B-C Lie groups. This is the case of the $Sp(N)$ groups: For example, the Dynkin diagram of $Sp(3)$ is given by mirroring the Dynkin diagram of $SU(6)$ as shown in Figure~\ref{DynkinSpSU} in Appendix \ref{sec_SUconi}. Translated to the fourfold, this means that the $\mathbb{P}^1$ associated with the dashed node is trivial under the monodromies corresponding to the mirroring. The corresponding fourfold exceptional divisor $E_{2N-1}$ is a $\mathbb{P}^1$ fibration over $\hat{D}$. On the other hand the other exceptional divisors $E_{2i-1}$ ($i=1,...,N-1$) are fibrations over $\hat{D}$ of two $\mathbb{P}^1$s, exchanged under the monodromy, once one goes around $\hat{D}$. These cycles can also be seen as a $\mathbb{P}^1$ fibration over the double cover $D$ of $\hat{D}$. Using the fact that the arithmetic genus of a $\mathbb{P}^1$ fibration over a manifold $M$ is the same as the arithmetic genus of $M$, one can 
derive the relations \eqref{chi0rel}.}

\section{Geometry of the $SU(N)$ resolved fourfold} \label{sec_SUconi}

In section~\ref{SpNgeom}, we discussed the features of establishing and resolving an $Sp(N)$ singularity. As it is known from the type IIB picture an $Sp(N)$ gauge group is broken to an $SU(N)\times U(1)$ gauge group by switching on a suitable gauge flux on the brane stack. So we could also directly impose an $SU(N)$ singularity in the F-theory picture even though there is an important difference to the breaking by flux that becomes apparent in the weak coupling limit.

Comparing the $SU(2N)$ and $Sp(N)$ Tate factorization in Table~\ref{SpSUSO} there is an additional factorization of $a_2$ which has a drastic consequence in the Sen limit. The Calabi-Yau hypersurface equation~\eqref{CY3TateSen} enforces a conifold singularity:
\begin{equation}
 0 = \xi^2 - (a_1^2 + w_i\,a_{2,1}) \equiv x_1 x_2-x_3x_4\,.
\label{SUNconifold}
\end{equation}
where on the RHS of eq.~\eqref{SUNconifold} we have used one of the standard parametrizations of the conifold~\cite{Candelas:1989js}, i.e.\ $x_1=\xi-a_1$, $x_2=\xi+a_1$, $x_3=w_i$ and $x_4=a_{2,1}$. Notice that this does not happen in the $Sp(N)$ case since $a_1^2 + a_2 =0$ cannot be brought into the form that parametrizes a conifold. 
Note that the appearance of this singularity does not depend on the gauge group rank, as it appears for all $SU(2N)$. Since it is not known how to resolve this conifold singularity there is no smooth transition between the general F-theory picture and the perturbative type IIB picture~\cite{Donagi:2009ra,Krause:2012yh,Collinucci:2012as}. In particular, we do not know if we can use the leading $\alpha'$ correction to the K\"ahler potential of~\cite{Becker:2002nn} which was derived in the smooth perturbative IIB picture.

The conifold sits at the point
\begin{equation}
\xi= a_1 = w_i = a_{2,1} = 0\ .
\end{equation}
We can check if this intersections exists in our example of an elliptically fibered fourfold over $B_3$ with an $SU(N)$ gauge group enforced over $\hat{D}_1$
\begin{equation}
\int_{B}\left[9 \hat{D}_1+2 \hat{D}_5\right]\wedge \hat{D}_1\wedge \left[18 \hat{D}_1+ 4 \hat{D}_5 - \hat{D}_1\right] = 22\,.\label{numconifolds}
\end{equation}
Hence, for our example we cannot avoid the conifold singularities.

\begin{figure}[h!]
\centering
\includegraphics[width= 0.65\linewidth]{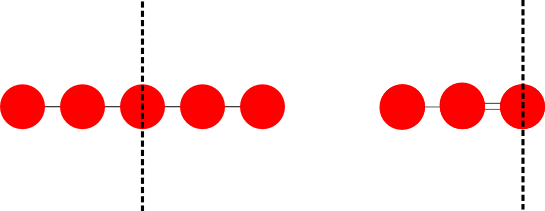}
\caption{Dynkin diagrams for $SU(6)$ (left) and $Sp(3)$ (right). The $Sp$ diagram can be obtained from the $SU$ diagram by mirroring the $SU$ diagram with respect to the dashed line. Only the blob which lies on the dashed line is invariant under this procedure.}
\label{DynkinSpSU}
\end{figure}

Ignoring the problematics of the Sen limit of $SU(N)$ gauge groups for the moment, we have attempted to calculate the arithmetic genus of the exceptional divisors in the resolved fourfold of an $SU(N)$ singularity imposed on a divisor $\hat{D}$. The blow-up procedure to obtain the resolved fourfold which is nicely discussed in~\cite{Krause:2012yh} for $SU(2),\, SU(3),\dots,SU(5)$ is more complicated than in the $Sp(N)$ case due to the following features:
\begin{itemize}
\item The scaling relations of the blown-up $\mathbb{P}_1$s labeled by exceptional divisors $e_i$ are e.g.\ for $SU(4)$ given by
\begin{equation}
\begin{array}{c|cccccc}
e_0 & X & Y & Z & e_1 & e_2 & e_3\\ 
\hline 
0 & 2 & 3 & 1 & 0 & 0 & 0\\ 
1 & 1 & 1 & 0 & -1 & 0 & 0\\ 
1 & 1 & 2 & 0 & 0 & 0 & -1\\ 
1 & 2 & 2 & 0 & 0 & -1 & 0\\ 
\end{array}\quad .
\label{SU4res}
\end{equation}
There is a `non-linear' scaling in $X$ and $Y$, i.e.\ the weights of $X$ and $Y$ of a line of weights are not proportional to the weights of the previous line which was the case in the $Sp(N)$ case eq.~\eqref{toric6fold}. As a consequence the Stanley Reisner ideal is not as easily derived as eq.~\eqref{SRSpN}.
\item The order in which the $e_i$'s are introduced does not always reproduce the Cartan matrix of $SU(N)$ in the double intersections of these exceptional divisors. This makes some relabeling inevitable which is reflected in the $SU(4)$ example eq.~\eqref{SU4res} by the $e_1,\,e_2,\,e_3$ columns not containing a diagonal matrix with entries $-1$.
\end{itemize}
However, the resolved fourfold including the double intersections of the $e_i$ were derived in~\cite{Krause:2012yh} up to $SU(5)$. We use the results of~\cite{Krause:2012yh} to calculate the arithmetic genus of the exceptional divisors and find
\begin{equation}
 \chi_0(e_i) = \frac{1}{12} \int_B \hat{D} \wedge \left[c_2(B)+ (\Kbar - \hat{D}) \wedge (\Kbar - 2 \hat{D}) \right]\quad \forall i < N \leq 5\,.\label{chi0ei}
\end{equation}
This result is the same as the $Sp(N)$ case only for the last introduced exceptional divisor $E_{2N-1}$, see eq.~\eqref{chi0vNapp} while the divisors $E_{2i-1}$ for $i<N$ obeyed a different formula. This is plausible as can be seen from the Dynkin diagrams of $Sp(N)$ and $SU(2N)$, see Figure~\ref{DynkinSpSU}.

Using eq.~\eqref{chi0ei} for our base $B$ and $\hat{D}=\hat{D}_1$ we see that all exceptional divisors have $\chi_0=1$ and hence could carry a non-perturbative superpotential from gaugino condensation. Even though we can only prove this for $SU(N)$ with $N\leq 5$ we suspect that this result will hold for arbitrary $SU(N)$ since the double intersections of the exceptional divisors are governed by the Cartan matrix which implies in particular that there is only an intersection of $e_i$ with the direct neighbors $e_{i-1},\, e_i$ and $e_{i+1}$. Hence, going to larger gauge group the arithmetic genus of say $e_1$ should not be affected by the newly added divisors $e_{N},\, e_{N-1},\dots$

\newpage
\bibliographystyle{JHEP.bst}
\bibliography{dSexample2}
\end{document}